\documentclass[aps,pra,preprint]{revtex4-1}
\usepackage{bm}
\usepackage[colorlinks=true,urlcolor=blue,linkcolor=blue,citecolor=blue]{hyperref}
\usepackage{color}
\usepackage{graphics}
\usepackage{graphicx}
\usepackage{epsfig}
\usepackage{amssymb}
\usepackage{amsmath}
\usepackage{hyperref}
\usepackage{physics}

\usepackage{array}                 

\usepackage{amsfonts}           

\begin{document}

\title{Non-equilibrium Coupling of a Quartz Resonator to Ions for Penning-Trap Fast Resonant Detection}

\author{Joaqu\'in~Berrocal$^{1}$} 
\author{Steffen~Lohse$^{2,3,4}$}
\author{Francisco~Dom\'inguez$^1$}
\author{Manuel J.~Guti\'errez$^{1}$}
\author{Francisco J.~Fern\'andez$^5$}
\author{Michael~Block$^{2,3,4}$}
\author{Juan J.~Garc\'ia-Ripoll$^6$}
\author{Daniel~Rodr\'iguez$^{1,7}$}\email[]{danielrodriguez@ugr.es}
\affiliation{
$^1$Departamento de F\'isica At\'omica, Molecular y Nuclear, Universidad de Granada, 18071 Granada, Spain \\
$^2$Department Chemie - Standort TRIGA, Johannes Gutenberg-Universität Mainz, D-55099, Mainz, Germany \\
$^3$GSI Helmholtzzentrum für Schwerionenforschung GmbH, D-64291, Darmstadt, Germany \\
$^4$Helmholtz-Institut Mainz, D-55099, Mainz, Germany \\
$^5$Departamento de Arquitectura y Tecnolog\'ia de Computadores, Universidad de Granada, 18071 Granada, Spain\\
$^6$Instituto de F\'isica Fundamental, Consejo Superior de  Investigaciones Cient\'ificas, Spain\\
$^7$Centro de Investigaci\'on en Tecnolog\'ias de la Informaci\'on y las Comunicaciones, Universidad de Granada, 18071 Granada, Spain
}

\date{\today}

\begin{abstract}
The coherent coupling between a quartz electro-mechanical resonator at room temperature and trapped ions in a 7-tesla Penning trap has been demonstrated for the first time. The signals arising from the coupling remain for integration times in the orders of seconds. From the measurements carried out, we demonstrate that the coupling allows detecting the reduced-cyclotron frequency ($\nu_+$) within times below 10~ms and with an improved resolution compared to conventional electronic detection schemes. A resolving power $\nu_+/\Delta \nu_+=2.4\times10^{7}$ has been reached in single measurements.  In this publication we present the first results, emphasizing the novel features of the quartz resonator as fast non-destructive ion-trap detector together with different ways to analyze the data and considering aspects like precision, resolution and sensitivity.
\end{abstract}
\maketitle
\section{Introduction}

Mass measurements on exotic nuclei with Penning traps widely rely on destructive detection techniques, like time-of-flight \cite{Koni1995} or phase-imaging ion-cyclotron-resonance \cite{Elis2013}, to measure the cyclotron frequency of the stored ions, after probing their eigenmotions and ejecting the ions out of the trap to a micro-channel plate detector. The time-of-flight (ion-cyclotron-resonance) technique has been successfully applied to extreme cases, remarkably, to measure the masses of several nobelium ($Z=102$) and lawrencium ($Z=103$) isotopes  \cite{Bloc2010,Mina2012,Bloc2019}. The phase-imaging (ion-cyclotron-resonance) technique has allowed recently extending these direct mass measurements to the first isotope in the group of superheavy elements $^{257}$Rf ($Z=104$). Further extending these measurements to determine the nuclear binding energies of elements in this region up to $Z=118$, has a strong interest due to the predicted enhancement of stability of these nuclei against fission (see e.g. \cite{Sobi1966}). $^{257}$Rf is produced in a fusion-evaporation reaction with a cross section of $\approx 15$~nb, which considering the efficiency of the SHIPTRAP Penning-trap facility \cite{Bloc2005}, results in a maximum rate of a single trapped ion every thirty minutes.  Such yield is further reduced (approximately a factor of three per atomic number) towards $Z=118$. This calls for the implementation of a new non-destructive Penning-trap detector with maximum efficiency and ultimate sensitivity.
In a Penning trap the ions are confined by the superposition of a high-homogeneous magnetic field and a quadrupolar electric potential \cite{Brow1986}. The magnetic field $B\vec e_z$ is in the axial direction and the motion of ions with a mass-to-charge ratio $m/q$ is the superposition of three independent motions, one parallel to $\vec B$ with a characteristic frequency 
\begin{equation}
\nu_z=\frac{1}{2\pi }\sqrt{\frac{q\;U}{m\;d^2}},
\label{eq_axialfrec}
\end{equation}
\noindent being $U$ the voltage that defines the confinement in the axial direction, and $d$ is a characteristic distance related to the distances from trap center to inner surfaces of the electrodes. The other two eigenmotions are in the radial plane with frequencies $\nu _+$ (modified-cyclotron) and $\nu _-$ (magnetron). These frequencies read
\begin{equation}
\nu_\pm=\frac{\nu_c}{2} \left( 1 \pm \sqrt{ 1 - 2 \left( \frac{\nu_z}{\nu_c} \right)^2 } \right),
\label{nu+1}
\end{equation}
with 
\begin{equation}
\nu_c=\frac{1}{2\pi}\frac{q}{m}B.
\label{nu_c}
\end{equation}

Very recently we have reported the first experimental results utilizing quartz resonators for measurements of $\nu_+$ of stored $^{40}$Ca$^+$ ions \cite{Lohs2019} using the Fourier-Transform Ion-Cyclotron-Resonance (FT-ICR) technique \cite{Comi1974,Comi1974_2}, and performed the first mass measurements using a quartz through a $\nu_+$ measurement on $^{206,207}$Pb$^+$ ions \cite{Lohs2020}, using FT-ICR combined with classical avoided crossing \cite{Corn1990}. In this publication we study the first coupling observed between trapped ions and a quartz resonator, even when the resonator is at room temperature. This coupling takes place within a short time window, where the resonator gains some energy from the driving field, applied to probe one of the eigenfrequencies ($\nu_+$) of the ions stored in the Penning trap.
\begin{figure}[t]
\centering\includegraphics[width=1\linewidth]{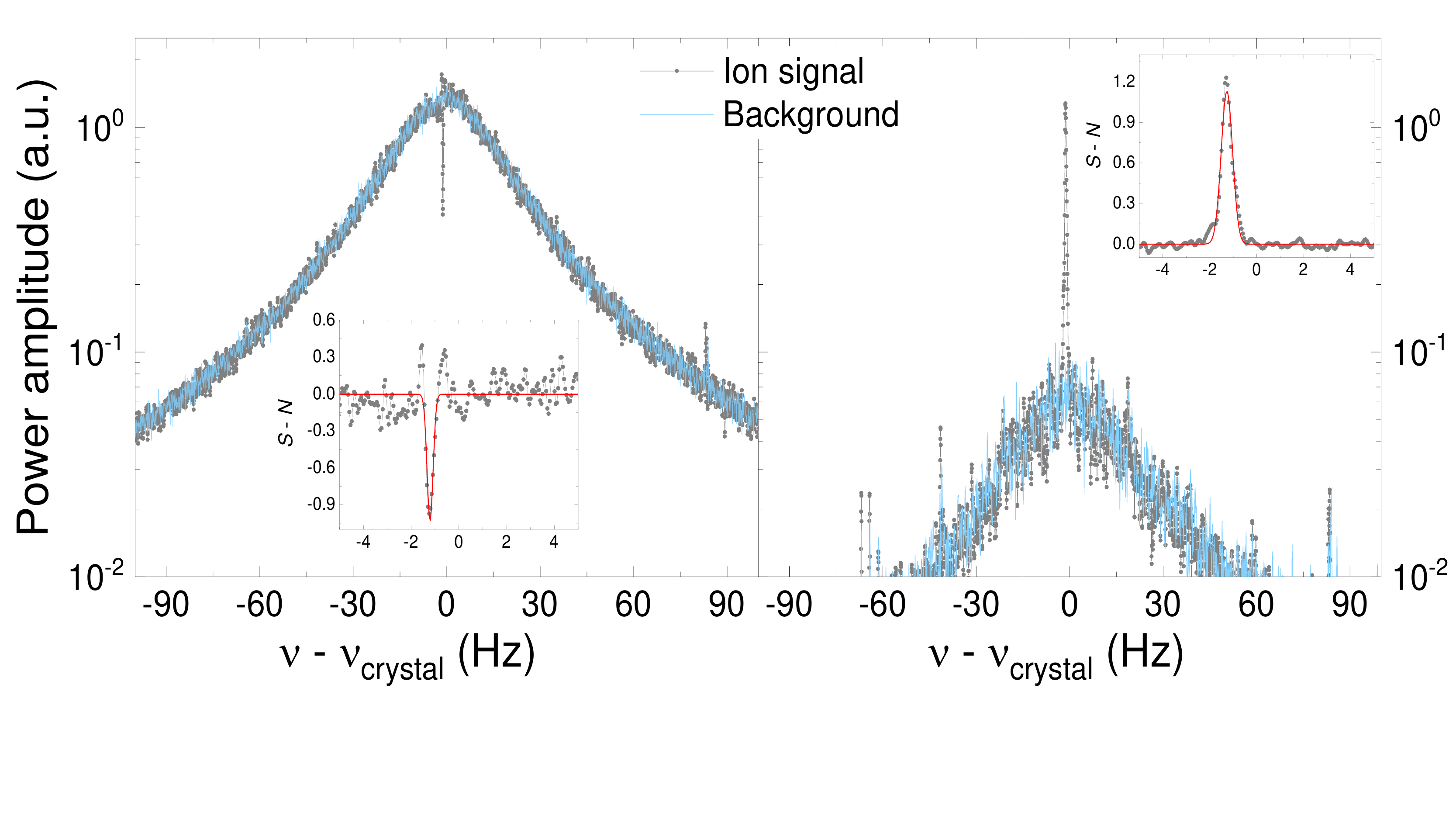}
\vspace{-2.3cm}
\caption{Left panel: Ions signal (dip) observed from the coupling between $^{40}$Ca$^+$ ions and the quartz crystal under non-equlibrium. Right panel: conventional FT-ICR ions signal (peak) observed from the current induced by trapped $^{40}$Ca$^+$ ions in the absence of  coherent interaction between the ions and the resonator. The electronic noise or background ($N$) is also shown. $\nu _{\scriptsize{\hbox{crystal}}}=2.68766141(4)$~MHz.  The insets zoom the regions around the dip and the peak, of the ions signal ($S$) minus background, and a Gaussian fit (red-solid line) yielding $\omega _+ =2\pi \times 2.68766019(4)$~MHz (left panel) and $\omega _+ =2\pi \times 2.68766013(4)$~MHz (right panel). The full width at half maximum of the dip is $295$~mHz, while that from the peak is $575$~mHz. \label{Figure1}}
\end{figure}
When the modified-cyclotron motion is coupled to the energized crystal, the ions' signal is observed as a dip on top of the noise (left panel of Fig.~\ref{Figure1}). When this coherence is lost during the thermalization of the crystal, the signal becomes a peak (right panel of Fig.~\ref{Figure1}) \cite{Lohs2019,Lohs2020}. Both signals were obtained from the Fourier transformation of the power dissipated on a fixed input resistance. Each set of data points is the average of 20 single measurements considering the same acquisition time-window of 4~seconds. The acquisition starts 2~ms before the driving field is stopped (left panel), and 90~ms after this has been stopped (right panel). Besides the advantage of providing an improvement in resolution, the ions signal can already be observed if the dip is only present for times in the order of milliseconds. 

Although state-of-the-art nuclear models predict half-lives in the order of hundreds of years for longest-lives isotopes of superheavy elements \cite{Giul2019}, a short measurement time is of prime interest for several short-lived (neutron-deficient) superheavy isotopes and other rare nuclides that can be produced at next-generation radioactive ion-beam facilites.  This concerns Penning trap facilities like for instance MATS at FAIR (Facility for Antiprotons and Ion Research) \cite{Rodr2010} and the single ion Penning trap SIPT at the Facility for Rare Isotope Beams (FRIB) \cite{Hama2019}, where high-rates from very short-lived nuclei are expected. The features of the coherent interaction between ions and crystal will be analyzed in this publication.

\section{Interaction between ions and quartz crystals}
\begin{figure}[t]
\centering\includegraphics[width=0.8\linewidth]{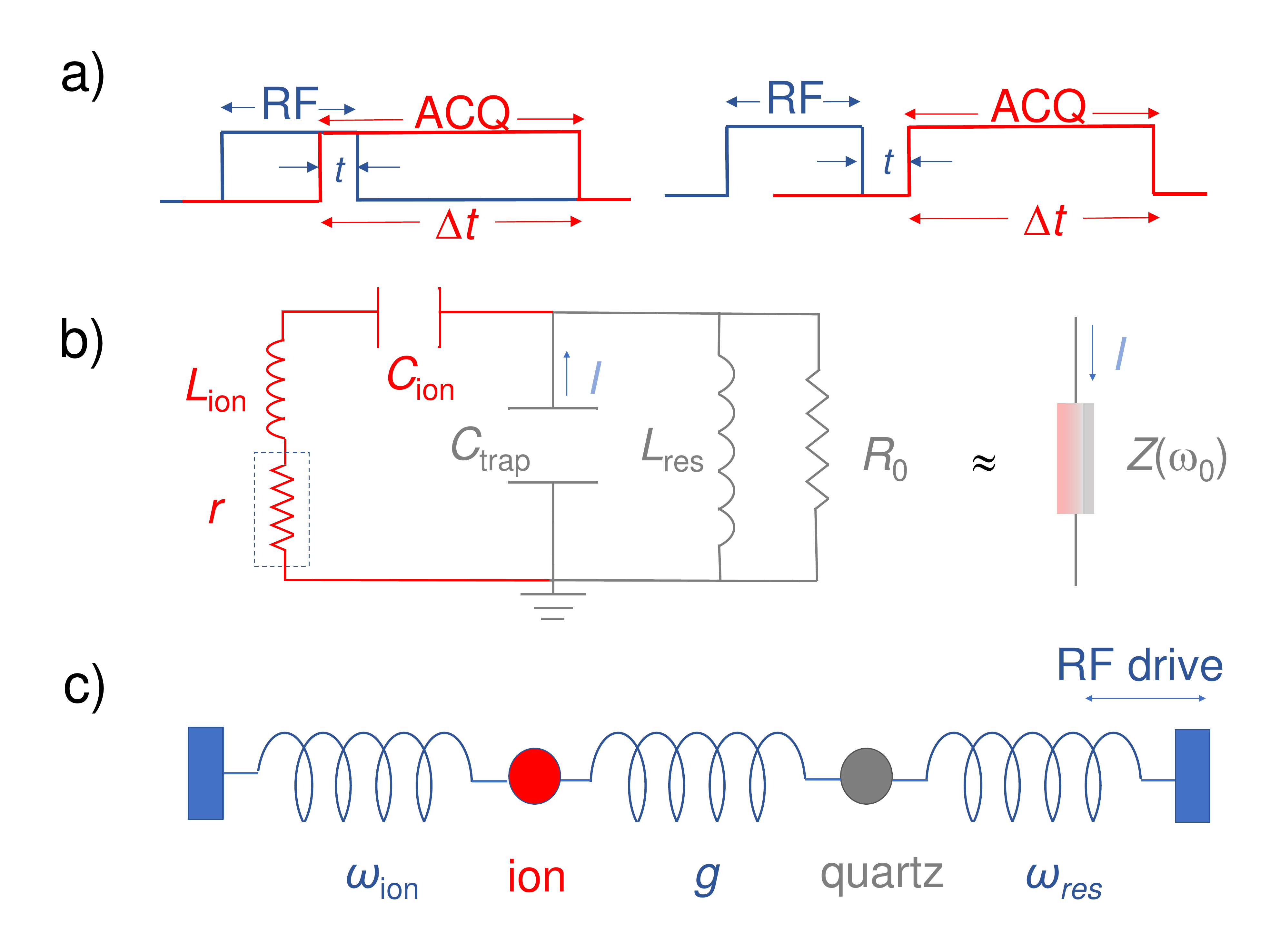}
\vspace{-0.8cm}
\caption{a) Time sequence indicating the delay ($t$) between excitation (RF drive) and acquisition (ACQ) of duration $\Delta t$. b) Schematics showing the equivalent circuit to describe the interaction between the ions and the resonator \cite{Wine1975}. The inductance $L_{\scriptsize{\hbox{ion}}}=mD_{\scriptsize{\hbox{eff}}}^{2}/(N_{\scriptsize{\hbox{ion}}}q^{2})$, depends on the mass $m$, the charge state $q$ of the ions, their number $N_{\scriptsize{\hbox{ion}}}$, and the effective distance $D_{\scriptsize{\hbox{eff}}}$ of the trap. The ion capacitance $C_{\scriptsize{\hbox{ion}}}=1/(\omega^{2}L_{\scriptsize{\hbox{ion}}})$ shows the dependency with the frequency. The resistance $r$ inside the dashed box is introduced as a dissipative term which prevents a full short-cut. c) Schematics showing the mechanical coupling considering the ion and resonator as driven harmonic oscillators. The constant $g$ describes the strength of the coupling. The RF drive will be represented by the function $\Omega (t)$. \label{sketch}}
\end{figure}

A quartz crystal has electrical and mechanical properties, which makes it different from a conventional resonator. Once the external driving field is stopped, the energy stored  in the crystal decays with a time constant, of about 8~ms. For a certain time window, this effect can also be observed in the Fourier transformation both for signals starting before of after the excitation. Besides the decay of the energy of the crystal, it is possible to observe from the Fourier transformation of the electrical power, the coherent coupling between the crystal and the ions as introduced in the left panel of Fig.~\ref{Figure1}. Two different but complementary views to model the ion-crystal interaction are shown in Fig.~\ref{sketch}: b) modeling the ion(s) by an equivalent circuit where the ion(s) is(are) in equilibrium with the trap and the resonator \cite{Wine1975}, and c) considering the coupling of two harmonic oscillators, the ion(s) and the resonator. Both models are equivalent for a certain phase of the ions' motion and the crystal oscillation. For the equivalent-circuit model, the electrical power through an input resistance $R_{\scriptsize{\hbox{input}}}$ is given by
\begin{equation}
P=(|Z(\omega)|\cdot I)^2/R_{\scriptsize{\hbox{input}}},
\end{equation}

\noindent where $I$ is the current induced by the trapped ions on one of the trap electrodes and $Z(\omega)$ is the impedance, which for the equivalent circuit (Fig.~\ref{sketch}b) is a complex number, whose square modulus can be written as 
\begin{equation}
\left|Z(\omega )\right|^{2}=\frac{R_0^{2}}{1+\left\lbrace Q_{\scriptsize{\hbox{res}}}f(\omega_{\scriptsize{\hbox{res}}})- \left[ Q_{\scriptsize{\hbox{ion}}}f(\omega_{\scriptsize{\hbox{ion}}})\right]^{-1} \right\rbrace ^{2}}, \label{eq:ion_lc} 
\end{equation}

\noindent where $f(\omega_{\scriptsize{\hbox{res}}})=\omega/\omega_{\scriptsize{\hbox{res}}}-\omega_{\scriptsize{\hbox{res}}}/\omega$, $f(\omega_{\scriptsize{\hbox{ion}}})=\omega/\omega_{\scriptsize{\hbox{ion}}}-\omega_{\scriptsize{\hbox{ion}}}/\omega$, $Q_{\scriptsize{\hbox{res}}}=R_0\sqrt{C_{\scriptsize{\hbox{trap}}}/L_{\scriptsize{\hbox{res}}}}$ and $Q_{\scriptsize{\hbox{ion}}}=R_0^{-1}\sqrt{L_{\scriptsize{\hbox{ion}}}/C_{\scriptsize{\hbox{ion}}}}$. $L_{\scriptsize{\hbox{ion}}}$ and $C_{\scriptsize{\hbox{ion}}}$ are defined in the caption of Fig.~\ref{sketch}. By introducing the resistance
 $r$ in Fig.~\ref{sketch}b, Eq.~(\ref{eq:ion_lc}) becomes 

\begin{equation}
\left|Z(\omega )\right|^{2}=\frac {R_0^2}{\left ( 1 + \frac{R_0}{r \left[ 1 + Q_{\scriptsize{\hbox{ion}}}^2f^2(\omega _{\scriptsize{\hbox{ion}}}) \right]} \right ) ^{2}
+\left ( Q_{\scriptsize{\hbox{res}}} f(\omega_{\scriptsize{\hbox{res}}}) + \frac{Q_{\scriptsize{\hbox{ion}}}R_0 f(\omega_{\scriptsize{\hbox{ion}}})}{r \left[ 1 + Q_{\scriptsize{\hbox{ion}}}^{2} f^2(\omega_{\scriptsize{\hbox{ion}}}) \right]} \right )^{2}}. \label{eq:ion_rlc} 
\end{equation} \\

The estimates of the equivalent circuit model are very useful for the spectroscopic study of the combined trap-ion system. However, if we wish to study the dynamical response of both elements, it is more convenient to use the equivalent model of two coupled oscillators (Fig.~\ref{sketch}c) with frequencies $\omega _{\scriptsize{\hbox{res}}}=2\pi\times \nu_{\scriptsize{\hbox{crystal}}}$ and $\omega _{\scriptsize{\hbox{ion}}}$, respectively, and coupling constant $g$. The Hamiltonian has the form
\begin{equation}
H= \hbar \omega_{\scriptsize{\hbox{ion}}} a^{\dagger}a + \hbar \omega _{\scriptsize{\hbox{res}}} b^{\dagger} b
+\hbar \Omega (t) b +\hbar \Omega (t)^{*}b^{\dagger} +(\hbar ga^{\dagger}b +\hbox{H.c.})
\end{equation}

\noindent where $ \Omega (t)$ is the driving field, which neglecting the counter rotating terms, it is given by $\Omega (t)=\Omega _0 \hbox{exp}(-i\omega _{\scriptsize{\hbox{RF}}}t)$. The coupling constant $g$ is a real or complex number that accounts for the strength and phase between the ions' motion and the oscillation of the resonator. In particular, the equivalent circuit model is recovered for a real and negative value of $g=-|g|<0$, but other models are possible, accounting for different phase relations between the ion and crystal oscillations. The system will evolve following a master equation of the form
\begin{equation}
\partial _t \rho = -\frac{i}{\hbar}[H,\rho]+L_1[\rho]+L_2[\rho]
\end{equation}

\noindent where $L_1$ and $L_2$ are Lindblad operators that model the thermalization of ions and crystal. At a finite temperature, these dissipators have the form
\begin{equation}
L[\rho] =\gamma (\overline{n} +1) \left(b\rho b^{\dagger}-\frac{1}{2}\lbrace b^{\dagger}b,\rho \rbrace \right ) +\gamma \overline {n} \left(b^{\dagger}\rho b-\frac{1}{2}\lbrace bb^{\dagger},\rho \rbrace \right ), \label{operator}
\end{equation}
\noindent where $\gamma $ represents the decay-time constant for both, ions ($\gamma _{\scriptsize{\hbox{ion}}}$) and resonator ($\gamma _{\scriptsize{\hbox{res}}}$), and $\overline {n}=\hbar \omega _{\scriptsize{\hbox{res}}}/k_BT$ is the average number of phonons at a given temperature, thus representing different states of the ions and the crystal. Equation~(\ref{operator}) describes the evolution in linear mode, so that the treatment of the problem is identical in the classical and in the quantum regime. The evolution of the Fock operators for the ion and for the resonator is given by

\begin{equation}
\frac{d\overline{a}}{dt}=\left (-i\omega -\frac{\gamma}{2}\right ) \overline {a} -ig(\overline {b^*}+\overline {b}), 
\end{equation}

\noindent and

\begin{equation}
\frac{d\overline{b}}{dt}=\left (-i\omega _{\scriptsize{\hbox{res}}} -\frac{\gamma _{\scriptsize{\hbox{res}}}}{2}\right ) \overline {b} -i\Omega (t) -ig(\overline {a^*}+\overline {a}),
\end{equation}
\noindent respectively. Since $g\ll\gamma _{\scriptsize{\hbox{res}}}\ll\omega _{\scriptsize{\hbox{res}}}$, it is possible to perform a perturbative expansion and apply some approximations. The response of the ion cloud can be very strong, even if the coupling is weak, due to the weaker friction. This response also has a different phase depending on the coupling constant, and on the detuning of the drive. This phase causes the coherent interference between the signal generated by the quartz oscillator and the ion's motion. The combined power of both signals, dissipated on a fixed input resistance is the square of the real part of $\overline {b}$, with $\overline {b}$ given by
\begin{equation}
\overline{b}(\omega)=\frac{1}{i(\omega _{\scriptsize{\hbox{res}}}-\omega)+\gamma _{\scriptsize{\hbox{res}}}/2}
\times \left (b_0+\frac{iga_0}{i(\omega _{\scriptsize{\hbox{ion}}}-\omega)+\gamma _{\scriptsize{\hbox{ion}}}/2}\right) \label{eq:b}
\end{equation}
being $a_0$ and $b_0$ the initial amplitudes of the oscillators representing the ions and the crystal, respectively. The other parameters are defined earlier. Note that in order to obtain Eq.~(\ref{eq:b}), we have considered that the acquisition time is infinite and we have neglected the loss of relative coherence that takes place at moderate times.

\section{Experimental setup and results}
The experiments reported here have been carried out with the TRAPSENSOR open-ring Penning trap \cite{Guti2019} and the quartz amplifier described in Ref.~\cite{Lohs2019}. Calcium atoms are evaporated from an oven and collimated through a 1-mm diameter hole to reach an area close to the center of the trap, where they are photoionized using two laser beams (Fig.~\ref{trap}), one tuned to be resonant with the $^1$S$_0\rightarrow ^1$P$_1$ transition in $^{40}$Ca ($\lambda \approx 422$~nm) and the other with a fixed wavelength ($\lambda \approx 375$~nm). The number of ions is regulated by the current applied to the oven. Figure~\ref{trap}a shows a cut of a three dimensional CAD drawing of the trap indicating the different electrodes and segments. The operation of a cold-head system with a copper structure attached and close to the trap, has made it possible to reach storage times in the order of 900~seconds (half-life). In order to optimize the trap performance, the potentials applied to the EC and CE electrodes are varied while the ring electrode is maintained at ground, and the FT-ICR ions signal is monitored until a maximum is reached. The frequency and the amplitude of the driving field remain the same in this procedure. The maximum signal indicates a better performance of the trap in terms of harmonicity, reducing the effect of higher-order terms in the expansion of the electrostatic potential. 
After the ions are created inside the trap, the photoionization lasers are blocked and two kinds of experiments are performed:
\begin{figure}[t]
\centering\includegraphics[width=0.9\linewidth]{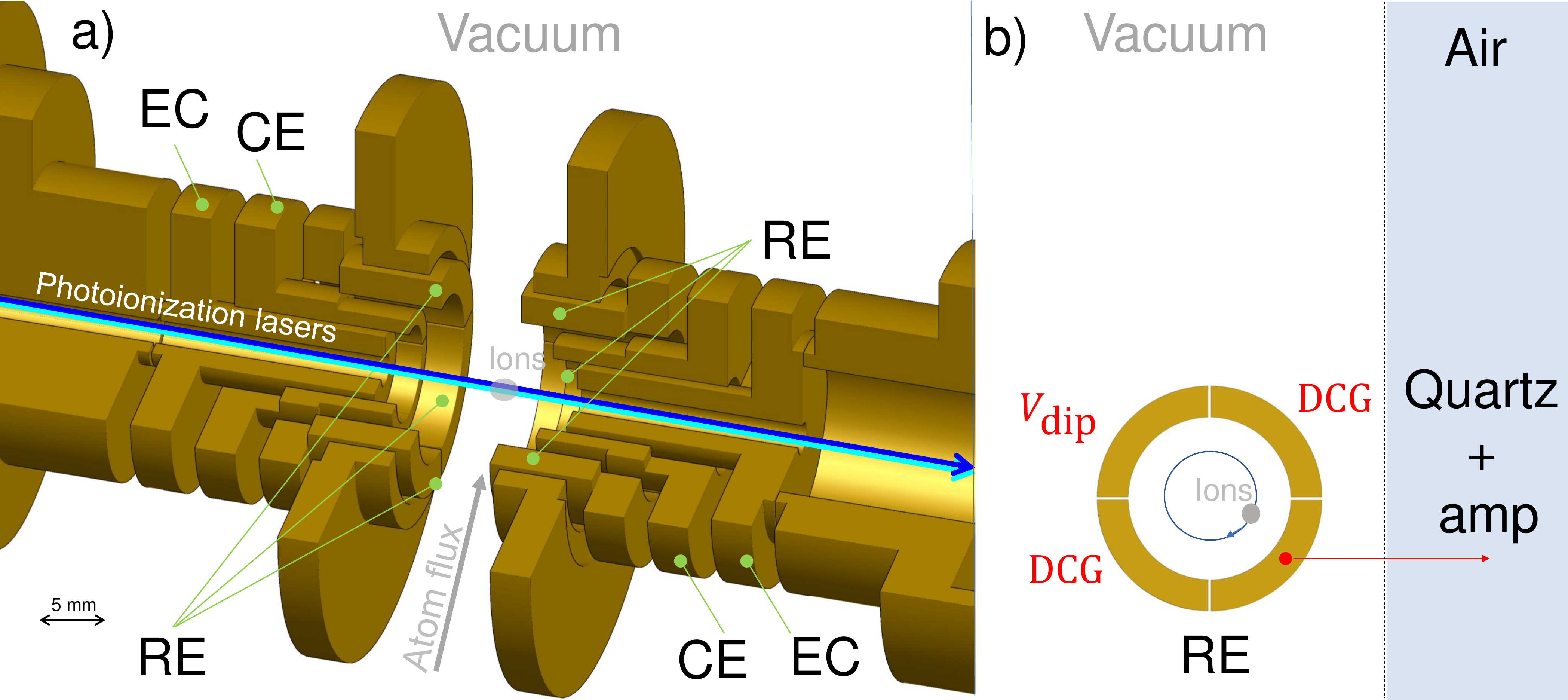}
\vspace{-0.2cm}
\caption{a) Cut from a CAD drawing showing the open-ring Penning trap used for the measurements \cite{Guti2019}. EC and CE stand for endcap and correction electrode, respectively. The ring electrode (RE) is four-fold segmented. b) Transverse cut of one of the ring electrodes. The external driving field is applied to one segment, two other segments are DC grounded (DCG) and the fourth one is connected to the quartz resonator, which is outside the vacuum vessel, through a wire \cite{Lohs2019}.  \label{trap}}
\end{figure}
\begin{enumerate}
\item The modified-cyclotron motion is directly probed by means of an external dipolar field in the radial direction applied to one of the segments of the ring electrode ($V_{\scriptsize{\hbox{dip}}}$ in Fig.~\ref{trap}b). The other ring electrodes are DC grounded. The field is applied for 400~ms, and data acquisition starts 2~ms before the excitation is stopped ($t=-2$~ms along the text). The ions remain in the trap during the time the acquisition is running ($4$~seconds) and then they are ejected from the trap and counted with a micro-channel plate (MCP) detector. 
\item The ions are laser cooled for 15 seconds. The components and operations added to the sketch in Fig.~\ref{trap}a to perform Doppler cooling are shown in Fig.~\ref{trap2}a. Twelve laser beams drive the transitions shown in Ref.~\cite{Guti2019} to perform Doppler cooling in both axial and radial directions. An external quadrupolar field at $\omega _{\scriptsize{\hbox{RF}}}=\omega _c$ is applied in the radial direction ($V_{\scriptsize{\hbox{quad}}}$ in Fig.~\ref{trap2}b) during the last 10~seconds of the cooling process, in order to reduce the ions' magnetron radii, before probing their modified-cyclotron motion. During the cooling process, the fluorescence photons are monitored with an Electron Multiplying Charge-Coupled Device (EMCCD) camera. After the cooling process, the dipolar field is also applied during 400~ms, and data acquisition starts 10~ms before the excitation is removed ($t=-10$~ms along the text). The ions remain in the trap during the 4~seconds of acquisition.
\end{enumerate}

\begin{figure}[t]
\centering\includegraphics[width=0.9\linewidth]{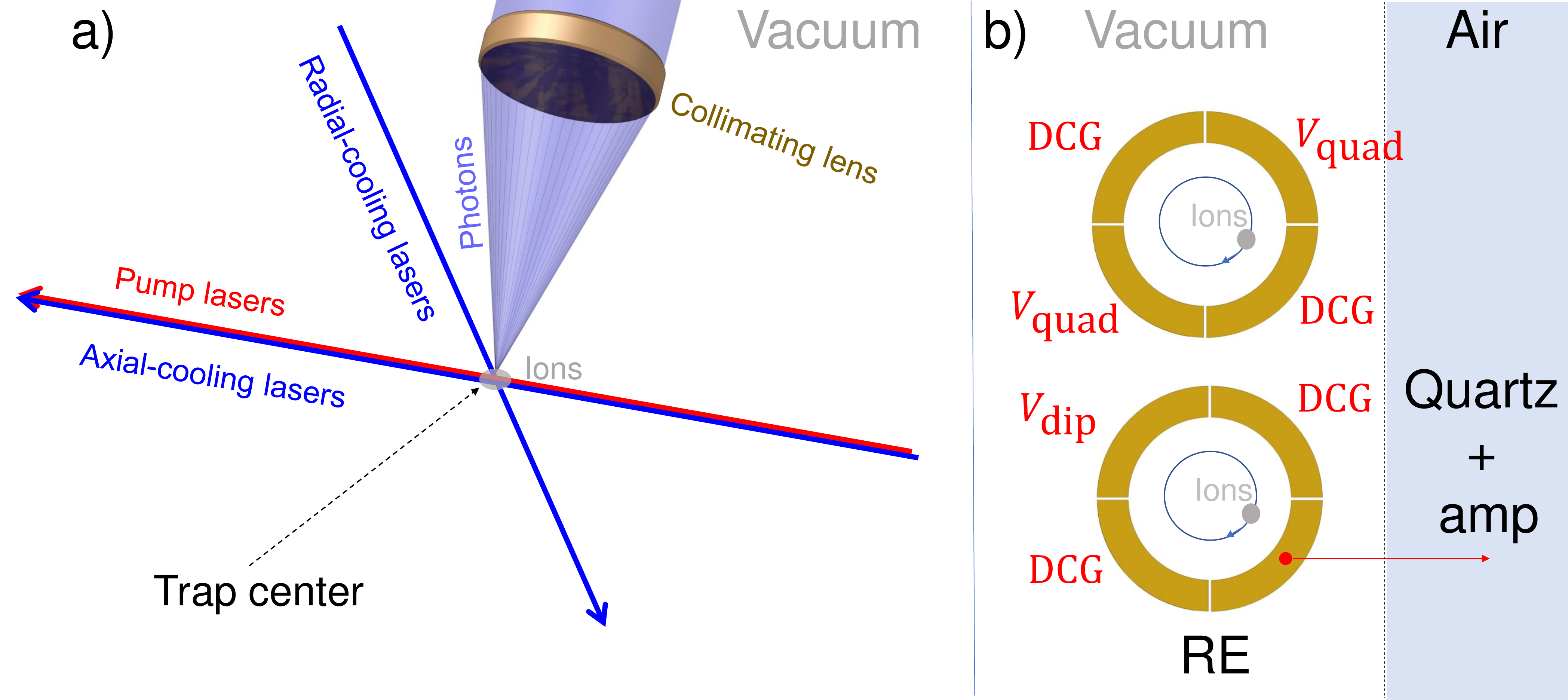}
\vspace{-0.3cm}
\caption{Elements and operations added to the system when laser cooling is applied. See text for details. \label{trap2}}
\end{figure}
\begin{figure*}[b]
\centering\includegraphics[width=1.0\linewidth]{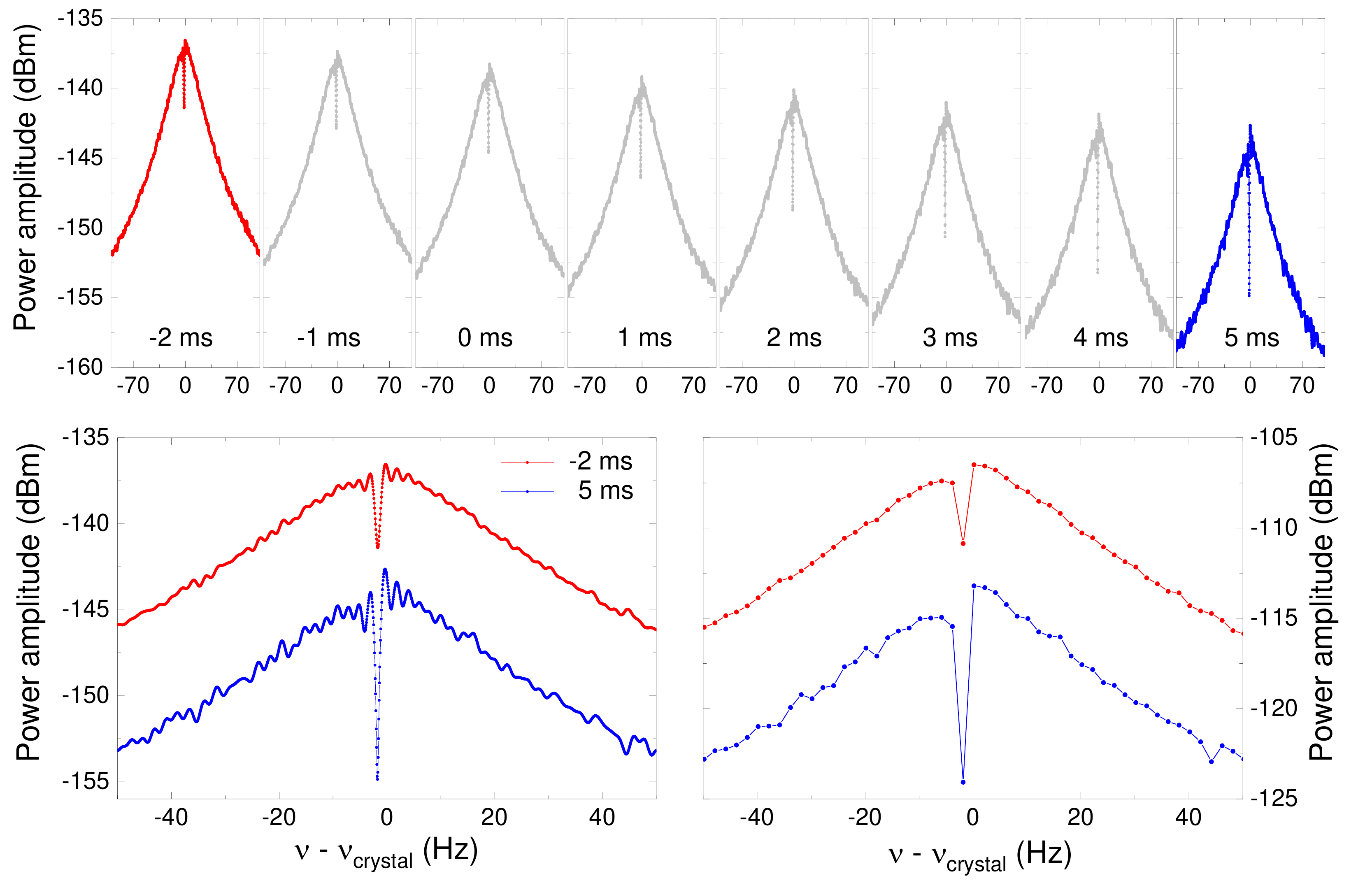}
\vspace{-0.9cm}
\caption{Top: Evolution of the ions signal versus time delay ($t$ in Fig.~\ref{sketch}a). The acquisition starts at $t=-2$~ms, and the excitation is stopped at $t=0$~ms. The time window considered for the Fourier transformation is 500~ms. Bottom: comparison between two signals considering different delays when zero-padding is applied (left) and when only the raw data are considered (right). \label{crystal_data_1}}
\end{figure*}
\begin{figure*}[t]
\centering\includegraphics[width=1.0\linewidth]{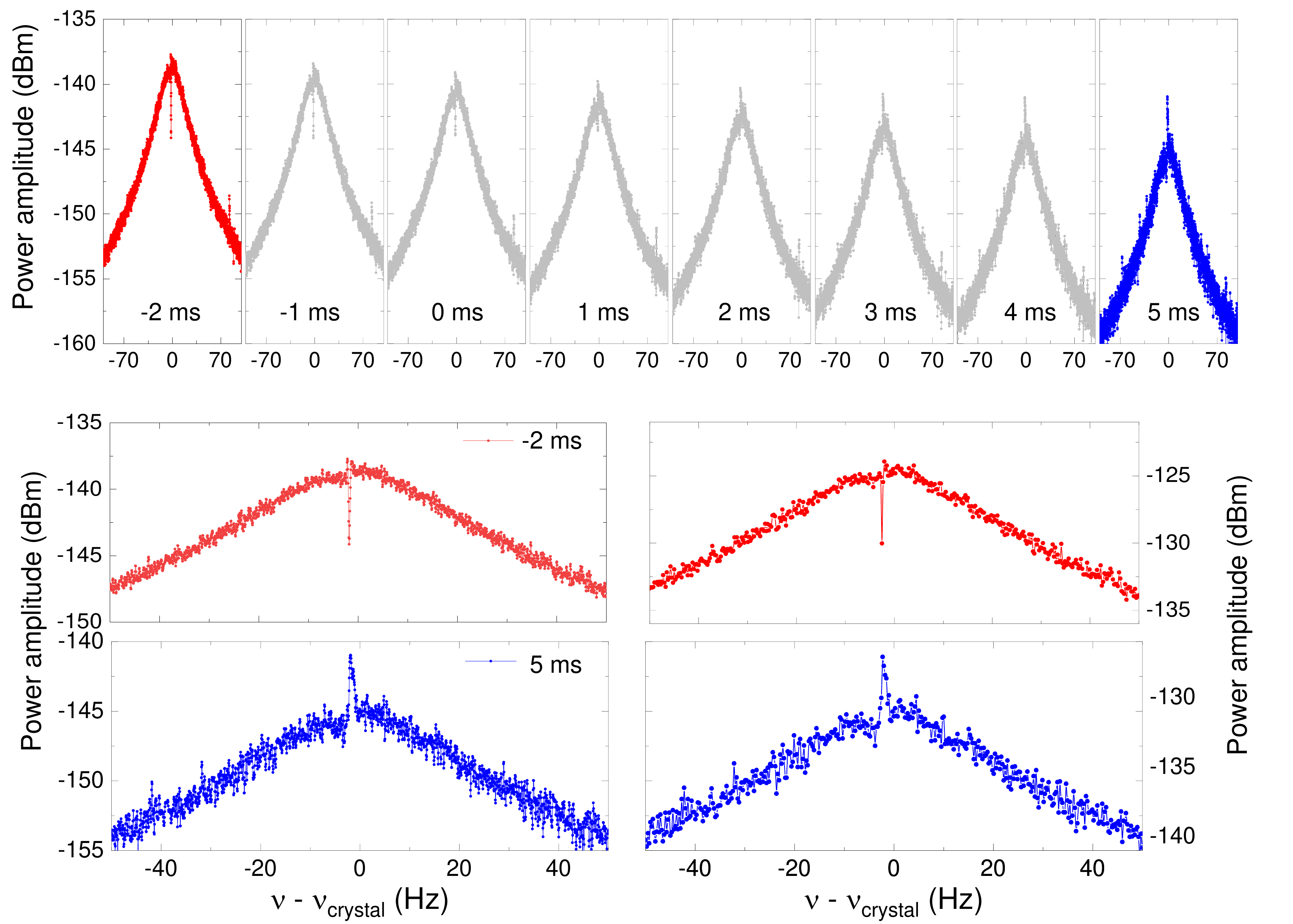}
\vspace{-0.9cm}
\caption{Top: Evolution of the ions signal versus time delay ($t$ in Fig.~\ref{sketch}a). At $t=0$~ms the excitation is stopped. The acquisition starts at $t=-2$~ms and the time window considered for the Fourier transformation is 4~seconds. Bottom: comparison between two signals considering different delays when zero-padding is applied (left) and when only the raw data are considered (right). \label{crystal_data_2}}
\end{figure*}

Measurements in the absence of laser cooling are presented in Figs.~\ref{crystal_data_1} and \ref{crystal_data_2} considering an acquisition time-window ($\Delta t$ in Fig.~\ref{sketch}a) of 0.5 and 4~seconds, respectively. The dipolar field has an amplitude of 2.5~mV$_{\scriptsize{\hbox{pp}}}$ and a frequency $\omega _{\scriptsize{\hbox{RF}}} = 2\pi\times 2.687659$~MHz close to $\omega _{\scriptsize{\hbox{res}}} \approx 2\pi\times 2.687661$~MHz and $\omega _{\scriptsize{\hbox{ion}}} \approx 2\pi \times 2.687660$~MHz. Zero padding was applied to increase the number of data points \cite{Juli2007} (see the comparison in both figures). A short acquisition time allows for better sensitivity (with respect to the ion number) since the coupling (originating the dip) is only present during a few milliseconds. The resolution in this case becomes worse. A longer time window increases the resolution but decreases the sensitivity in terms of ion number and requires more stable conditions in the experiment.

The data have been analyzed and fitted using the model described in Fig.~\ref{sketch} with the parameters of the equivalent-circuit representation, considering the full acquisition time-window of 4~seconds. The number of ions is around a few thousand. Smaller number of ions have been used in experiments where laser cooling has been performed before applying the driving field. In such scenario, the cooling in the radial direction, needs an external quadrupolar field (Fig.~\ref{trap2}b) with $\omega _{\scriptsize{\hbox{RF}}}=\omega _c$ in order to reduce the ions' magnetron radii.
\begin{figure}[t]
\centering\includegraphics[width=1.0\linewidth]{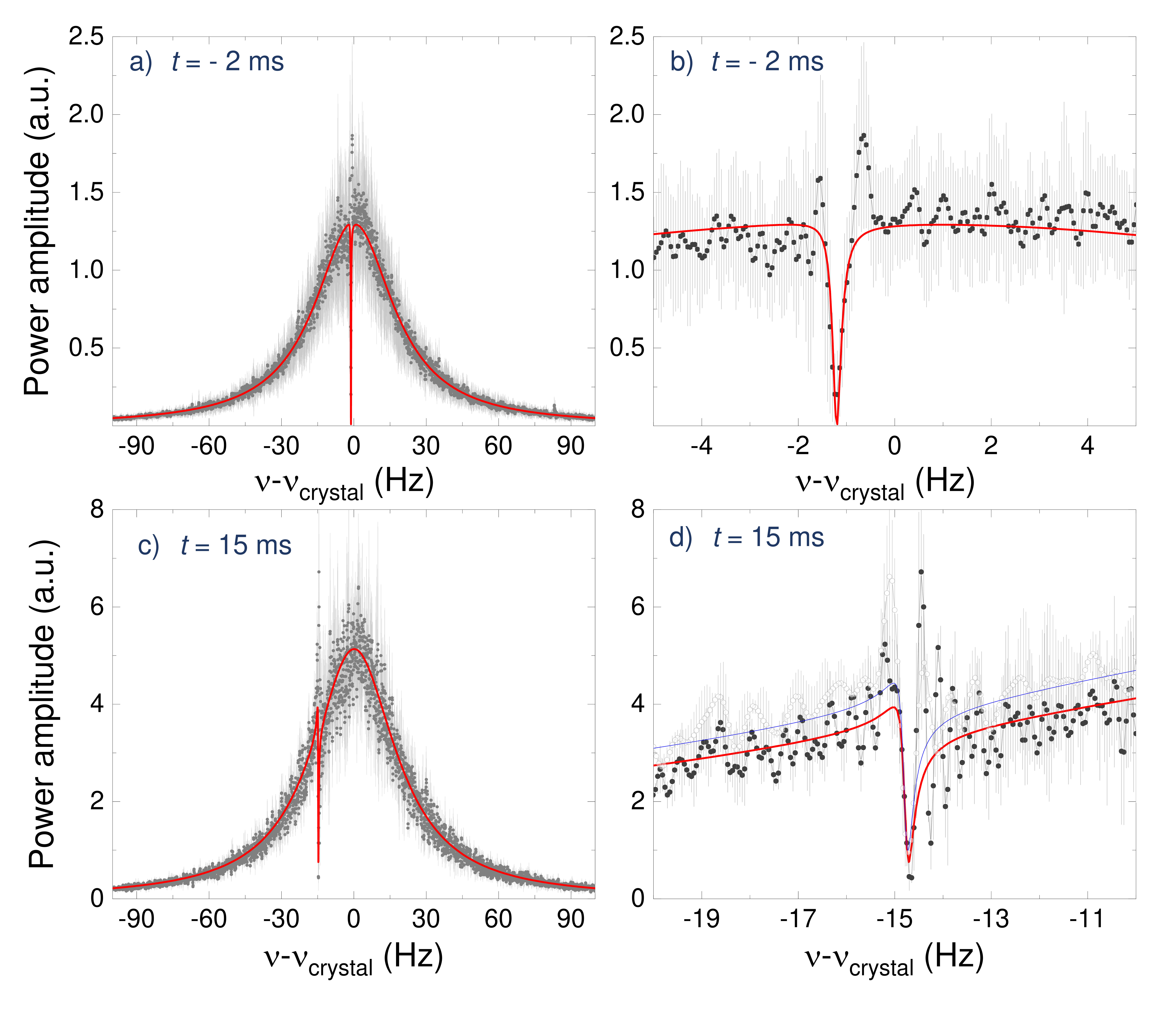}
\vspace{-1.5cm}
\caption{Fit to the data points using the function in Eq.~(\ref{eq:ion_rlc}).  Upper panel: the frequency of the crystal is very close to that of the ions. Lower panel:  The frequency of the crystal is about 15~Hz higher than $\nu _+$. b) and d) zoom plots around $\nu _{\scriptsize{\hbox{+}}}$ (dip) of the data in a) and b), respectively. The open circles and the blue-solid line in d) correspond to the same measurement and fit curve (Eq.~(\ref{eq:ion_rlc})) but considering $\Delta t=2$~seconds (and zero padding). 13 and 3 measurements were used to obtain the data points in the upper and in the lower panel, respectively. Further details are given in the text. \label{data_1}}
\end{figure}
Figure~\ref{data_1} shows the results from the fits using Eq.~(\ref{eq:ion_rlc}) considering data from $t=-2$~ms (upper panels) and from $t=15$~ms (lower panels). Laser cooling was applied to carry out the measurements presented in the lower panel. In these cases, the amplitude of the driving field is in the range from 30 to 40~mV$_{\scriptsize{\hbox{pp}}}$. In order to perform the fit using the multi-parameter function (Eqs.~(\ref{eq:ion_lc}) or (\ref{eq:ion_rlc}) if a full short-cut is not observed), $D_{\scriptsize{\hbox{eff}}}=40$~mm and $R_0=700$~k$\Omega$ \cite{Lohs2019}. The values of $\omega _{\scriptsize{\hbox{res}}}$ and $Q_{\scriptsize{\hbox{res}}}$ are obtained from a Lorentzian fit to the ions signal, yielding $\omega_{\scriptsize{\hbox{res}}}=2\pi\times 2.68766141(4)$~MHz and $Q_{\scriptsize{\hbox{res}}}\approx 67,000$ for the results shown in the upper panel. $r$ and $\omega_+$ are free parameters. Larger weighing (a factor $\approx 10^4$ compared to 1) has been given to data points around the dip and to a certain range of data points at both sides from the dip along the Lorentzian profile. Under these conditions one obtains from the fit $\omega_+=2\pi\times 2.687660(83)$~MHz, and $r\approx 470(5)$~k$\Omega$ for  about $6000$~ions. The latter number is obtained from the calibration with the MCP detector. Due to the large number of ions used in the measurements presented in Fig.~\ref{data_1}a and zoomed in b, the resolution for the 4-second acquisition time-window was sufficient to see the dependence of the width of the fit curve on $N_{\scriptsize{\hbox{ion}}}$. For the measurement presented in Fig.~\ref{data_1}c and zoomed in d, the amplitude of the driving field was $40$~mV$_{\scriptsize{\hbox{pp}}}$, $Q_{\scriptsize{\hbox{res}}}=64,000$ and $\nu _{\scriptsize{\hbox{crystal}}}\approx \nu_+ +15$~Hz. The value of $\nu_{\scriptsize{\hbox{crystal}}}=2.6876693(3)$~MHz for these measurements is different compared to $\nu_{\scriptsize{\hbox{crystal}}}$ for the measurements in the upper panel. This is due to a different positioning of the crystal in the printed circuit board of the amplifier. In order to perform these measurements, the voltages applied to the EC and CE electrodes were slightly tuned to shift the frequency closer to the resonance frequency of the crystal maintaining the harmonicity of the electrostatic potential well. Under these conditions, one obtains from the fit using Eq.~(\ref{eq:ion_rlc}), $\omega_+=2\pi\times 2.68765(30)$~MHz. Figure.~\ref{data_1}d shows also data points (open circles) and the fit curve (blue-solid line) when considering $\Delta t=2$~seconds. As mentioned earlier, though the resolution becomes worse, the ion sensitivity improves. From the number of photons registered with a photomultiplier tube, $N_{\scriptsize{\hbox{ion}}}$ for the results in the lower panel (red-solid line) is smaller, although it can not be extracted from the fit. In order to see the dependence on the ion number, one would need a higher resolution and thus a longer acquisition time-window.

Although the models in Fig.~\ref{sketch} allow explaining to some extent the expected interaction between the crystal and the ions, the conditions of the experiment were not sufficient to determine the reduced-cyclotron frequency precisely even when several measurements are averaged to reduce the fluctuations in each data point. These fluctuations lead to a jitter of the signal that washes out over different measurements. For this purpose we have developed a second procedure that consists in subtracting the background (electronic noise) from the ions' signal and obtain the reduced-cyclotron frequency from a Gaussian fit (inset of Fig.~\ref{Figure1}). By performing such a fit, for example, to the signals at $t=-2$~ms in Figs.~\ref{crystal_data_1} and \ref{crystal_data_2}, the reduced-cyclotron frequency values differ in about $80$~mHz, corresponding to a relative frequency shift of $3\times 10^{-8}$. The full width at half maximum ($FWHM)$ is 1~Hz for the results shown in Fig.~\ref{crystal_data_1} and 295~mHz for those in Fig.~\ref{crystal_data_2}. 

\begin{figure}[t]
\centering\includegraphics[width=0.7\linewidth]{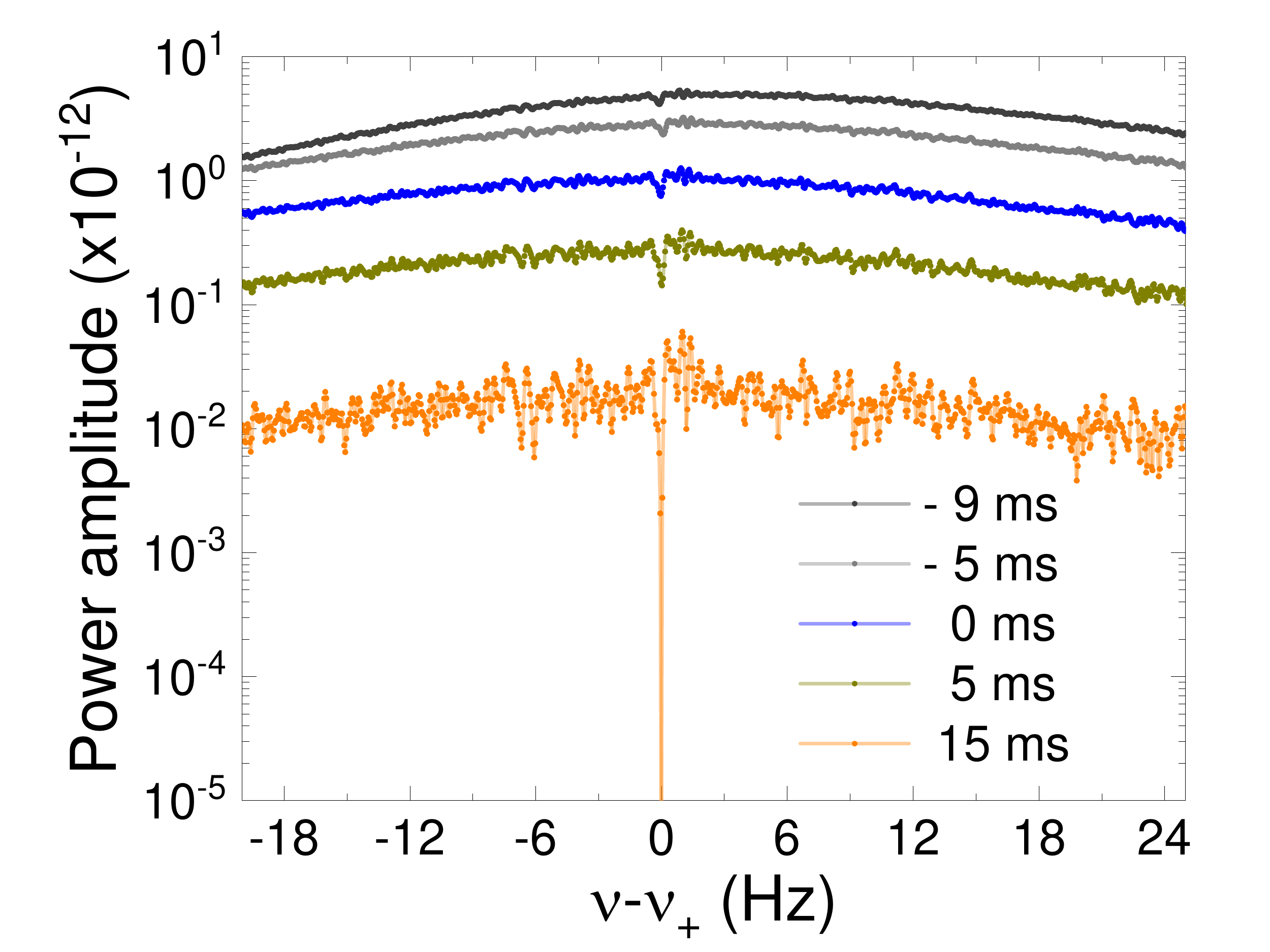}
\vspace{-0.4cm}
\caption{Evolution of the signal from an individual measurement for different time delays ($t$ in Fig.~\ref{sketch}a) of the acquisition relative to the end of the excitation. Laser cooling is performed before applying the driving field. $\nu_{\scriptsize{\hbox{RF}}}=2.687681$~MHz, $\nu_+ = 2.687669$~MHz and $\nu_{\scriptsize{\hbox{crystal}}}$ varies from $2.687675$~MHz  at $t=-10$~ms to $2.687670$~MHz at $t=15$~ms. \label{data_laser_2}}
\end{figure}

\begin{figure}[t]
\centering\includegraphics[width=0.8\linewidth]{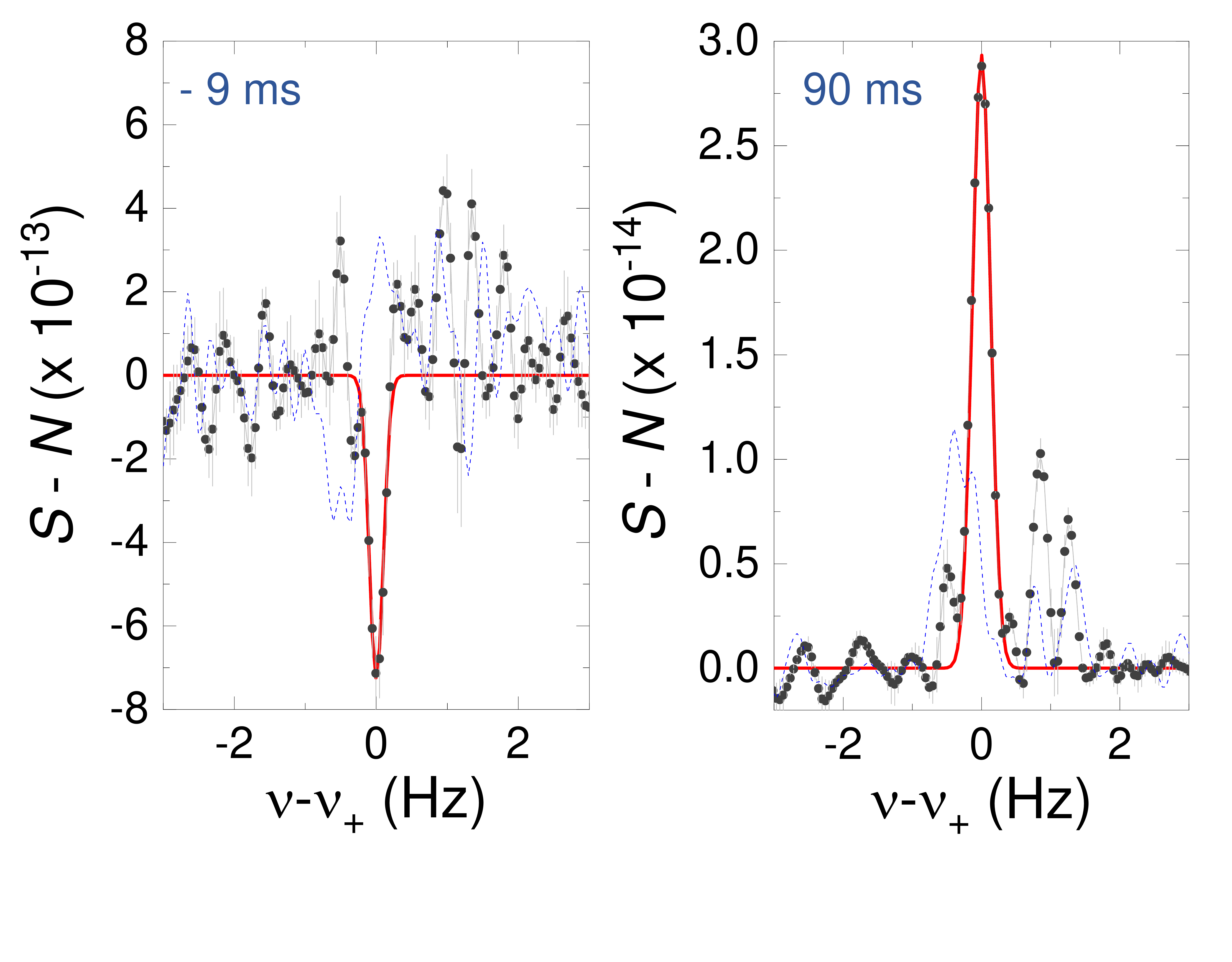}
\vspace{-1.5cm}
\caption{Comparison of peak and dip effective signals, i.e. ions signals ($S$) minus electronic noise ($N$), considering different time delays on starting the acquisition time-window ($t$ in Fig.~\ref{sketch}a). The data are from the same measurement shown in Fig.~\ref{data_laser_2}. Besides the data points and the fit (red-solid line), the blue-dashed line in both panels represents an ion cloud signal from another single measurement .}\label{data_2}
\end{figure}

\begin{figure}[t]
\centering\includegraphics[width=1.0\linewidth]{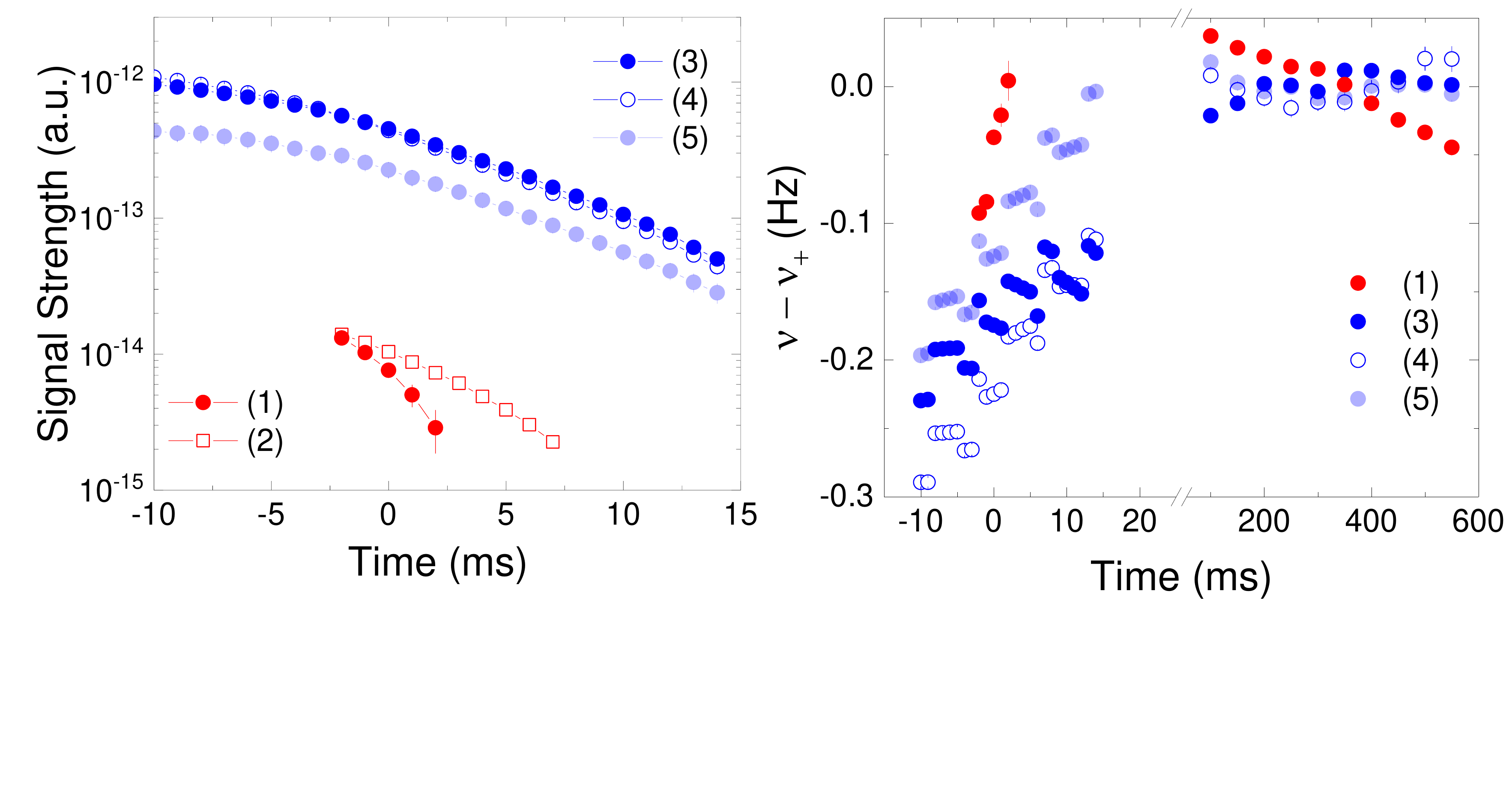}
\vspace{-3.0cm}
\caption{Comparison of the results from Gaussian fits to effective ion signals  from measurements under different conditions. (1) and (2) are results from the same measurement, but the latter considering an acquisition time-window of 0.5~seconds (Fig.~\ref{crystal_data_1}). $\nu_+$ on the right is taken as the mean value of the center of the peaks distributions resulting from subtracting the electronic noise to the ions signal for each measurement ($t>100$~ms). See text for details. \label{decay}}
\end{figure}

Figure~\ref{data_laser_2} shows the evolution of the ions signal of a single measurement when laser cooling is performed before the excitation. One can observe that the frequency of the crystal changes towards the equilibrium value, while the value of $\nu_+$ does not vary. The coupling of the crystal to the external field is stronger compared to the coupling of the ions. Figure~\ref{data_2} (left) shows the Gaussian fit to the effective ions signal at $t=-9$~ms in Fig.~\ref{data_laser_2}, and the Gaussian fit when $t=90$~ms (right). The result from one measurement cycle (out of twenty) is shown. The ions signal from the other measurement cycles (see e.g. blue-dashed line in the same figure) are not clearly visible and we assign this to some instabilities in the system. Important features from this treatment of the data are presented in Fig.~\ref{decay} and in Tabs.~(\ref{summary}) and (\ref{summary2}).

\begin{table}[t] 
\caption{Experimental parameters and precision of the measurements presented in Fig.~\ref{decay} when $\Delta t =4$~seconds. $\sigma (\nu _+^d)$ is the standard deviation of the $\nu_+$ value from the measurements when $t<15$~ms (dips) and $\sigma (\nu _+^p)$ is the one obtained from the measurements when $t>100$~ms (peaks). (*) rounded values. }
\vspace{0.3cm}
\centering
 \renewcommand{\arraystretch}{1.3}
\setlength{\tabcolsep}{2.2mm}

	\begin{tabular}{cccccccc}
			\hline\hline
Measurement &  Cooling  & $N_{\scriptsize{\hbox{ion}}}$&  $V_{\scriptsize{\hbox{RF}}}$&  $\nu_+ -\nu_{\scriptsize{\hbox{RF}}}$& $\nu_+ -\nu_{\scriptsize{\hbox{crystal}}}$&  $\sigma (\nu _+^d)$& $\sigma (\nu _+^p)$  \\
(Fig.~\ref{decay})&          &                                               & (mV$_{\scriptsize{\hbox{pp}}}$)&                                                     (Hz)&                        (Hz)&                      (mHz)& (mHz)\\

\hline \hline
			(1)&  No&  $\approx 6000$&  2.5&  1.14& -1.37& 50&  28  \\
                         (3)&  Yes&  $\approx 750$&  40&  -14*& -15*& 33&  10  \\
                         (4)&  Yes&  $\approx 560$&  35&  -12*& -0.73&  57&  13  \\
                         (5)&  Yes&  $\approx 280$&  30&  -15*& -5*& 58&   8 \\
                  	\hline \hline 
	\end{tabular} \label{summary}
\end{table}

\begin{table}[t] 
\caption{Resolving power in $\nu _+$ from the measurements presented in Fig.~\ref{decay}.}
\vspace{0.3cm}
\centering
 \renewcommand{\arraystretch}{1.3}
\setlength{\tabcolsep}{2.2mm}

	\begin{tabular}{ccccccc}
			\hline\hline
Measurement &  Cooling  & $N_{\scriptsize{\hbox{ion}}}$ & $\overline {FWHM}_d$ & $\overline {FWHM}_p$&   $\left(\frac{\nu_+}{\Delta \nu_+}\right)_d$&    $\left(\frac{\nu_+}{\Delta \nu_+}\right )_p$ \\
(Fig.~\ref{decay})&  & & (mHz)& (mHz)& & \\

\hline \hline
		       (1)&  No&   $\approx 6000$&  332&  605& $9.8\times 10^{6}$& $4.4\times 10^{6}$\\
                        (3)&  Yes&   $\approx 750$&  129&  657& $2.2\times 10^{7}$&  $4.1\times 10^{6}$\\
                        (4)&  Yes&   $\approx 560$&  243&  340& $1.2\times 10^{7}$&  $7.9\times 10^{6}$\\
                        (5)&  Yes&  $\approx 280$&  116&  213& $2.4\times 10^{7}$& $1.3\times 10^{7}$ \\
                                          	\hline \hline 
	\end{tabular} \label{summary2}
\end{table}

Figure~\ref{decay} shows on the left panel the ratio between the amplitude  and the width of the Gaussian distribution resulting from the fit of the effective ions signal (see e.g. Fig.~\ref{data_2}), as a function of the delay-time $t$. (1) and (2) are obtained from the ions signals presented in Fig.~\ref{crystal_data_1} and \ref{crystal_data_2}, respectively. (3), (4) (partly shown in Fig.~\ref{data_laser_2} and~\ref{data_2}) and (5) are single measurements where laser cooling was performed. Due to the cooling, the amplitude of the driving field is increased by more than one order of magnitude (Tab.~(\ref{summary})) and this increases the sensitivity of the effective ion signal. It also increases the time the coherent interaction between the ions and the crystal can be observed after the driving field is stopped. In (3), (4) and (5) the radiofrequency of the driving field was more than 10~Hz above $\nu_+$, which results in less power broadening provided $\nu_{\scriptsize{\hbox{crystal}}}$ differs also a few Hz from $\nu_+$ (Tab.~\ref{summary2}). As shown in Tab.~(\ref{summary}), the standard deviation of $\sigma (\nu _+^d)$ is below 100~mHz. Although the right panel of Fig.~\ref{decay} shows an increasing trend of $\nu_+$, the difference between minimum and maximum is small so that one can consider $\nu_+$ as the mean value of those resulting from the fit when $t<15$~ms. The standard deviations of the frequency values extracted from peaks ($t>100$~ms) are smaller and differ slightly so that from any of the measurement (dip or peak) one can determine $\nu_+$ with a relative uncertainty in the order of $10^{-7}$ if the ions remain in the trap while the driving field is applied. Table~(\ref{summary2}) complements the results shown in Tab.~(\ref{summary}) comparing the mean $FWHM$  and the resolving power in $\nu_+$ for the dip and the peak distributions. The dip is better resolved than the peak. 

\section{Conclusions and Outlook}

In this work we have shown for the first time the coupling of trapped ions to a quartz resonator under non-equilibrium, even at room temperature and installed outside the vacuum vessel. This kind of measurement introduces novel features to Penning trap mass spectroscopy, compared to conventional FT-ICR measurements carried out with the same amplifier \cite{Lohs2019,Lohs2020}. The coherent coupling yields a resolution in the measurement of the reduced-cyclotron frequency, which is at least a factor of $2$ better, and only requires detection times in the order of milliseconds. This makes this technique a promising one for experiments on exotic nuclei. Regarding sensitivity, the number of trapped ions has been gradually reduced in subsequent experiments where laser cooling has been performed before probing the ions. Furthermore, we have been able to re-use the same ions in different subsequent measurements by performing in each cycle, laser cooling after excitation and data acquisition. The minimum number of ions for the measurements presented here is $\approx 280$. Performing measurements on a lower number of ions would require longer acquisition times and better stability of the power supply compared to the one utilized to supply the voltages to the EC and CE electrodes.

If instead of a power measurement we perform a full quadrature measurement of the combined oscillators signals, we could get access to a characterization of the relative phase between the ions and the quartz resonator. The combination of both gives rise to a very precise Fano-like profile, that would increase our sensitivity to the ion's motional frequency within the same time-window. Besides the importance of the results for fast detection in nuclear physics experiments, the characteristics of the quartz, might open up new possibilities if the ion is cooled to the ground state of motion in the Penning trap. These experimental results illustrate the coupling between an ion cloud and a high-quality radiofrequency resonator in the weak coupling regime. While these experiments are performed in semiclassical conditions, they open the way towards exploring non-equilibrium hybrid quantum systems and quantum correlations using higher-quality resonators in cryogenic environments \cite{Kotl2017}.

\section*{Acknowledgement}
We acknowledge support from the Spanish MCINN through the projects FPA2015-67694-P, and PID2019-104093GB-I00/AEI/10.01339/501100011033, from the Spanish Ministry of Education through PhD fellowship FPU17/02596, and from the University of Granada "Plan propio - Programa de Intensificaci\'on de la Investigaci\'on", project PP2017-PRI.I-04 and "Laboratorios Singulares 2020". The construction of the facility was supported by the European Research Council (contract number 278648-TRAPSENSOR), the above-mentioned project FPA2015-67694-P, project FPA2012-32076, and the infrastructure projects UNGR10-1E-501, UNGR13-1E-1830 and EQC2018-005130-P (MICINN/FEDER/\linebreak Universidad de Granada), and the infrastructure projects IE-5713 and IE2017-5513 (Junta de Andaluc\'ia-FEDER). MB and SL acknowledge financial support by the German BMBF in the project 05P18UMFN1. JJGR acknowledges support from CSIC Technology Platform PTI-001 and  CAM PRICYT project QUITEMAD+CM S2013-ICE2801.\\

\section*{References}

\begin{thebibliography}{20}%
\makeatletter
\providecommand \@ifxundefined [1]{%
 \@ifx{#1\undefined}
}%
\providecommand \@ifnum [1]{%
 \ifnum #1\expandafter \@firstoftwo
 \else \expandafter \@secondoftwo
 \fi
}%
\providecommand \@ifx [1]{%
 \ifx #1\expandafter \@firstoftwo
 \else \expandafter \@secondoftwo
 \fi
}%
\providecommand \natexlab [1]{#1}%
\providecommand \enquote  [1]{``#1''}%
\providecommand \bibnamefont  [1]{#1}%
\providecommand \bibfnamefont [1]{#1}%
\providecommand \citenamefont [1]{#1}%
\providecommand \href@noop [0]{\@secondoftwo}%
\providecommand \href [0]{\begingroup \@sanitize@url \@href}%
\providecommand \@href[1]{\@@startlink{#1}\@@href}%
\providecommand \@@href[1]{\endgroup#1\@@endlink}%
\providecommand \@sanitize@url [0]{\catcode `\\12\catcode `\$12\catcode
  `\&12\catcode `\#12\catcode `\^12\catcode `\_12\catcode `\%12\relax}%
\providecommand \@@startlink[1]{}%
\providecommand \@@endlink[0]{}%
\providecommand \url  [0]{\begingroup\@sanitize@url \@url }%
\providecommand \@url [1]{\endgroup\@href {#1}{\urlprefix }}%
\providecommand \urlprefix  [0]{URL }%
\providecommand \Eprint [0]{\href }%
\providecommand \doibase [0]{http://dx.doi.org/}%
\providecommand \selectlanguage [0]{\@gobble}%
\providecommand \bibinfo  [0]{\@secondoftwo}%
\providecommand \bibfield  [0]{\@secondoftwo}%
\providecommand \translation [1]{[#1]}%
\providecommand \BibitemOpen [0]{}%
\providecommand \bibitemStop [0]{}%
\providecommand \bibitemNoStop [0]{.\EOS\space}%
\providecommand \EOS [0]{\spacefactor3000\relax}%
\providecommand \BibitemShut  [1]{\csname bibitem#1\endcsname}%
\let\auto@bib@innerbib\@empty
\bibitem [{\citenamefont {K{\"{o}}nig}\ \emph {et~al.}(1995)\citenamefont
  {K{\"{o}}nig}, \citenamefont {Bollen}, \citenamefont {Kluge}, \citenamefont
  {Otto},\ and\ \citenamefont {Szerypo}}]{Koni1995}%
  \BibitemOpen
  \bibfield  {author} {\bibinfo {author} {\bibfnamefont {M.}~\bibnamefont
  {K{\"{o}}nig}}, \bibinfo {author} {\bibfnamefont {G.}~\bibnamefont {Bollen}},
  \bibinfo {author} {\bibfnamefont {H.-J.}\ \bibnamefont {Kluge}}, \bibinfo
  {author} {\bibfnamefont {T.}~\bibnamefont {Otto}}, \ and\ \bibinfo {author}
  {\bibfnamefont {J.}~\bibnamefont {Szerypo}},\ }\href {\doibase
  10.1016/0168-1176(95)04146-C} {\bibfield  {journal} {\bibinfo  {journal}
  {Int. J. Mass Spectrom. Ion Processes}\ }\textbf {\bibinfo {volume} {142}},\
  \bibinfo {pages} {95 } (\bibinfo {year} {1995})}\BibitemShut {NoStop}%
\bibitem [{\citenamefont {Eliseev}\ \emph {et~al.}(2013)\citenamefont
  {Eliseev}, \citenamefont {Blaum}, \citenamefont {Block}, \citenamefont
  {Droese}, \citenamefont {Goncharov}, \citenamefont {Minaya~Ramirez},
  \citenamefont {Nesterenko}, \citenamefont {Novikov},\ and\ \citenamefont
  {Schweikhard}}]{Elis2013}%
  \BibitemOpen
  \bibfield  {author} {\bibinfo {author} {\bibfnamefont {S.}~\bibnamefont
  {Eliseev}}, \bibinfo {author} {\bibfnamefont {K.}~\bibnamefont {Blaum}},
  \bibinfo {author} {\bibfnamefont {M.}~\bibnamefont {Block}}, \bibinfo
  {author} {\bibfnamefont {C.}~\bibnamefont {Droese}}, \bibinfo {author}
  {\bibfnamefont {M.}~\bibnamefont {Goncharov}}, \bibinfo {author}
  {\bibfnamefont {E.}~\bibnamefont {Minaya~Ramirez}}, \bibinfo {author}
  {\bibfnamefont {D.~A.}\ \bibnamefont {Nesterenko}}, \bibinfo {author}
  {\bibfnamefont {Y.~N.}\ \bibnamefont {Novikov}}, \ and\ \bibinfo {author}
  {\bibfnamefont {L.}~\bibnamefont {Schweikhard}},\ }\href {\doibase
  10.1103/PhysRevLett.110.082501} {\bibfield  {journal} {\bibinfo  {journal}
  {Phys. Rev. Lett.}\ }\textbf {\bibinfo {volume} {110}},\ \bibinfo {pages}
  {082501} (\bibinfo {year} {2013})}\BibitemShut {NoStop}%
\bibitem [{\citenamefont {Block}\ \emph {et~al.}(2010)\citenamefont {Block},
  \citenamefont {Ackermann}, \citenamefont {Blaum}, \citenamefont {Droese},
  \citenamefont {Dworschak}, \citenamefont {Eliseev}, \citenamefont
  {Fleckenstein}, \citenamefont {Haettner}, \citenamefont {Herfurth},
  \citenamefont {Hessberger}, \citenamefont {Hofmann}, \citenamefont
  {Ketelaer}, \citenamefont {Ketter}, \citenamefont {Kluge}, \citenamefont
  {Marx}, \citenamefont {Mazzocco}, \citenamefont {Novikov}, \citenamefont
  {Plass}, \citenamefont {Popeko}, \citenamefont {Rahaman}, \citenamefont
  {{Rodr{\'i}guez}}, \citenamefont {Scheidenberger}, \citenamefont
  {Schweikhard}, \citenamefont {Thirolf}, \citenamefont {Vorobyev},\ and\
  \citenamefont {Weber}}]{Bloc2010}%
  \BibitemOpen
  \bibfield  {author} {\bibinfo {author} {\bibfnamefont {M.}~\bibnamefont
  {Block}}, \bibinfo {author} {\bibfnamefont {D.}~\bibnamefont {Ackermann}},
  \bibinfo {author} {\bibfnamefont {K.}~\bibnamefont {Blaum}}, \bibinfo
  {author} {\bibfnamefont {C.}~\bibnamefont {Droese}}, \bibinfo {author}
  {\bibfnamefont {M.}~\bibnamefont {Dworschak}}, \bibinfo {author}
  {\bibfnamefont {S.}~\bibnamefont {Eliseev}}, \bibinfo {author} {\bibfnamefont
  {T.}~\bibnamefont {Fleckenstein}}, \bibinfo {author} {\bibfnamefont
  {E.}~\bibnamefont {Haettner}}, \bibinfo {author} {\bibfnamefont
  {F.}~\bibnamefont {Herfurth}}, \bibinfo {author} {\bibfnamefont {F.~P.}\
  \bibnamefont {Hessberger}}, \bibinfo {author} {\bibfnamefont
  {S.}~\bibnamefont {Hofmann}}, \bibinfo {author} {\bibfnamefont
  {J.}~\bibnamefont {Ketelaer}}, \bibinfo {author} {\bibfnamefont
  {J.}~\bibnamefont {Ketter}}, \bibinfo {author} {\bibfnamefont {H.-J.}\
  \bibnamefont {Kluge}}, \bibinfo {author} {\bibfnamefont {G.}~\bibnamefont
  {Marx}}, \bibinfo {author} {\bibfnamefont {M.}~\bibnamefont {Mazzocco}},
  \bibinfo {author} {\bibfnamefont {Y.~N.}\ \bibnamefont {Novikov}}, \bibinfo
  {author} {\bibfnamefont {W.~R.}\ \bibnamefont {Plass}}, \bibinfo {author}
  {\bibfnamefont {A.}~\bibnamefont {Popeko}}, \bibinfo {author} {\bibfnamefont
  {S.}~\bibnamefont {Rahaman}}, \bibinfo {author} {\bibfnamefont
  {D.}~\bibnamefont {{Rodr{\'i}guez}}}, \bibinfo {author} {\bibfnamefont
  {C.}~\bibnamefont {Scheidenberger}}, \bibinfo {author} {\bibfnamefont
  {L.}~\bibnamefont {Schweikhard}}, \bibinfo {author} {\bibfnamefont {P.~G.}\
  \bibnamefont {Thirolf}}, \bibinfo {author} {\bibfnamefont {G.~K.}\
  \bibnamefont {Vorobyev}}, \ and\ \bibinfo {author} {\bibfnamefont
  {C.}~\bibnamefont {Weber}},\ }\href {\doibase 10.1038/nature08774} {\bibfield
   {journal} {\bibinfo  {journal} {Nature}\ }\textbf {\bibinfo {volume}
  {463}},\ \bibinfo {pages} {785} (\bibinfo {year} {2010})}\BibitemShut
  {NoStop}%
\bibitem [{\citenamefont {Minaya~Ramirez}\ \emph {et~al.}(2012)\citenamefont
  {Minaya~Ramirez}, \citenamefont {Ackermann}, \citenamefont {Blaum},
  \citenamefont {Block}, \citenamefont {Droese}, \citenamefont
  {D{\"{u}}llmann}, \citenamefont {Dworschak}, \citenamefont {Eibach},
  \citenamefont {Eliseev}, \citenamefont {Haettner}, \citenamefont {Herfurth},
  \citenamefont {He{\ss}berger}, \citenamefont {Hofmann}, \citenamefont
  {Ketelaer}, \citenamefont {Marx}, \citenamefont {Mazzocco}, \citenamefont
  {Nesterenko}, \citenamefont {Novikov}, \citenamefont {Pla{\ss}},
  \citenamefont {Rodr{\'\i}guez}, \citenamefont {Scheidenberger}, \citenamefont
  {Schweikhard}, \citenamefont {Thirolf},\ and\ \citenamefont
  {Weber}}]{Mina2012}%
  \BibitemOpen
  \bibfield  {author} {\bibinfo {author} {\bibfnamefont {E.}~\bibnamefont
  {Minaya~Ramirez}}, \bibinfo {author} {\bibfnamefont {D.}~\bibnamefont
  {Ackermann}}, \bibinfo {author} {\bibfnamefont {K.}~\bibnamefont {Blaum}},
  \bibinfo {author} {\bibfnamefont {M.}~\bibnamefont {Block}}, \bibinfo
  {author} {\bibfnamefont {C.}~\bibnamefont {Droese}}, \bibinfo {author}
  {\bibfnamefont {C.~E.}\ \bibnamefont {D{\"{u}}llmann}}, \bibinfo {author}
  {\bibfnamefont {M.}~\bibnamefont {Dworschak}}, \bibinfo {author}
  {\bibfnamefont {M.}~\bibnamefont {Eibach}}, \bibinfo {author} {\bibfnamefont
  {S.}~\bibnamefont {Eliseev}}, \bibinfo {author} {\bibfnamefont
  {E.}~\bibnamefont {Haettner}}, \bibinfo {author} {\bibfnamefont
  {F.}~\bibnamefont {Herfurth}}, \bibinfo {author} {\bibfnamefont {F.~P.}\
  \bibnamefont {He{\ss}berger}}, \bibinfo {author} {\bibfnamefont
  {S.}~\bibnamefont {Hofmann}}, \bibinfo {author} {\bibfnamefont
  {J.}~\bibnamefont {Ketelaer}}, \bibinfo {author} {\bibfnamefont
  {G.}~\bibnamefont {Marx}}, \bibinfo {author} {\bibfnamefont {M.}~\bibnamefont
  {Mazzocco}}, \bibinfo {author} {\bibfnamefont {D.}~\bibnamefont
  {Nesterenko}}, \bibinfo {author} {\bibfnamefont {Y.~N.}\ \bibnamefont
  {Novikov}}, \bibinfo {author} {\bibfnamefont {W.~R.}\ \bibnamefont
  {Pla{\ss}}}, \bibinfo {author} {\bibfnamefont {D.}~\bibnamefont
  {Rodr{\'\i}guez}}, \bibinfo {author} {\bibfnamefont {C.}~\bibnamefont
  {Scheidenberger}}, \bibinfo {author} {\bibfnamefont {L.}~\bibnamefont
  {Schweikhard}}, \bibinfo {author} {\bibfnamefont {P.~G.}\ \bibnamefont
  {Thirolf}}, \ and\ \bibinfo {author} {\bibfnamefont {C.}~\bibnamefont
  {Weber}},\ }\href {\doibase 10.1126/science.1225636} {\bibfield  {journal}
  {\bibinfo  {journal} {Science}\ }\textbf {\bibinfo {volume} {337}},\ \bibinfo
  {pages} {1207} (\bibinfo {year} {2012})}\BibitemShut {NoStop}%
\bibitem [{\citenamefont {Block}(2019)}]{Bloc2019}%
  \BibitemOpen
  \bibfield  {author} {\bibinfo {author} {\bibfnamefont {M.}~\bibnamefont
  {Block}},\ }\href {https://doi.org/10.1515/ract-2019-0002} {\bibfield
  {journal} {\bibinfo  {journal} {Radiochimica Acta}\ }\textbf {\bibinfo
  {volume} {20190002}} (\bibinfo {year} {2019})}\BibitemShut {NoStop}%
\bibitem [{\citenamefont {Sobiczewski}\ \emph {et~al.}(1966)\citenamefont
  {Sobiczewski}, \citenamefont {Gareev},\ and\ \citenamefont
  {Kalinkin}}]{Sobi1966}%
  \BibitemOpen
  \bibfield  {author} {\bibinfo {author} {\bibfnamefont {A.}~\bibnamefont
  {Sobiczewski}}, \bibinfo {author} {\bibfnamefont {F.}~\bibnamefont {Gareev}},
  \ and\ \bibinfo {author} {\bibfnamefont {B.}~\bibnamefont {Kalinkin}},\
  }\href {https://www.sciencedirect.com/science/article/pii/0031916366912431}
  {\bibfield  {journal} {\bibinfo  {journal} {Phys.\ Lett.}\ }\textbf {\bibinfo
  {volume} {22}},\ \bibinfo {pages} {500} (\bibinfo {year} {1966})}\BibitemShut
  {NoStop}%
\bibitem [{\citenamefont {Block}\ \emph {et~al.}(2005)\citenamefont {Block},
  \citenamefont {Ackermann}, \citenamefont {Beck}, \citenamefont {Blaum},
  \citenamefont {Breitenfeldt}, \citenamefont {Chauduri}, \citenamefont
  {Doemer}, \citenamefont {Eliseev}, \citenamefont {Habs}, \citenamefont
  {Heinz}, \citenamefont {Herfurth}, \citenamefont {He{\ss}berger},
  \citenamefont {Hofmann}, \citenamefont {Geissel}, \citenamefont {Kluge},
  \citenamefont {Kolhinen}, \citenamefont {Marx}, \citenamefont {Neumayr},
  \citenamefont {Mukherjee}, \citenamefont {Petrick}, \citenamefont {Plass},
  \citenamefont {Quint}, \citenamefont {Rahaman}, \citenamefont {Rauth},
  \citenamefont {Rodr{\'i}guez}, \citenamefont {Scheidenberger}, \citenamefont
  {Schweikhard}, \citenamefont {Suhonen}, \citenamefont {Thirolf},
  \citenamefont {Wang}, \citenamefont {Weber},\ and\ \citenamefont {{the
  SHIPTRAP Collaboration}}}]{Bloc2005}%
  \BibitemOpen
  \bibfield  {author} {\bibinfo {author} {\bibfnamefont {M.}~\bibnamefont
  {Block}}, \bibinfo {author} {\bibfnamefont {D.}~\bibnamefont {Ackermann}},
  \bibinfo {author} {\bibfnamefont {D.}~\bibnamefont {Beck}}, \bibinfo {author}
  {\bibfnamefont {K.}~\bibnamefont {Blaum}}, \bibinfo {author} {\bibfnamefont
  {M.}~\bibnamefont {Breitenfeldt}}, \bibinfo {author} {\bibfnamefont
  {A.}~\bibnamefont {Chauduri}}, \bibinfo {author} {\bibfnamefont
  {A.}~\bibnamefont {Doemer}}, \bibinfo {author} {\bibfnamefont
  {S.}~\bibnamefont {Eliseev}}, \bibinfo {author} {\bibfnamefont
  {D.}~\bibnamefont {Habs}}, \bibinfo {author} {\bibfnamefont {S.}~\bibnamefont
  {Heinz}}, \bibinfo {author} {\bibfnamefont {F.}~\bibnamefont {Herfurth}},
  \bibinfo {author} {\bibfnamefont {F.~P.}\ \bibnamefont {He{\ss}berger}},
  \bibinfo {author} {\bibfnamefont {S.}~\bibnamefont {Hofmann}}, \bibinfo
  {author} {\bibfnamefont {H.}~\bibnamefont {Geissel}}, \bibinfo {author}
  {\bibfnamefont {H.~J.}\ \bibnamefont {Kluge}}, \bibinfo {author}
  {\bibfnamefont {V.}~\bibnamefont {Kolhinen}}, \bibinfo {author}
  {\bibfnamefont {G.}~\bibnamefont {Marx}}, \bibinfo {author} {\bibfnamefont
  {J.~B.}\ \bibnamefont {Neumayr}}, \bibinfo {author} {\bibfnamefont
  {M.}~\bibnamefont {Mukherjee}}, \bibinfo {author} {\bibfnamefont
  {M.}~\bibnamefont {Petrick}}, \bibinfo {author} {\bibfnamefont
  {W.}~\bibnamefont {Plass}}, \bibinfo {author} {\bibfnamefont
  {W.}~\bibnamefont {Quint}}, \bibinfo {author} {\bibfnamefont
  {S.}~\bibnamefont {Rahaman}}, \bibinfo {author} {\bibfnamefont
  {C.}~\bibnamefont {Rauth}}, \bibinfo {author} {\bibfnamefont
  {D.}~\bibnamefont {Rodr{\'i}guez}}, \bibinfo {author} {\bibfnamefont
  {C.}~\bibnamefont {Scheidenberger}}, \bibinfo {author} {\bibfnamefont
  {L.}~\bibnamefont {Schweikhard}}, \bibinfo {author} {\bibfnamefont
  {M.}~\bibnamefont {Suhonen}}, \bibinfo {author} {\bibfnamefont {P.~G.}\
  \bibnamefont {Thirolf}}, \bibinfo {author} {\bibfnamefont {Z.}~\bibnamefont
  {Wang}}, \bibinfo {author} {\bibfnamefont {C.}~\bibnamefont {Weber}}, \ and\
  \bibinfo {author} {\bibnamefont {{the SHIPTRAP Collaboration}}},\ }\href
  {\doibase 10.1140/epjad/i2005-06-013-5} {\bibfield  {journal} {\bibinfo
  {journal} {Eur. Phys. J. A}\ }\textbf {\bibinfo {volume} {25}},\ \bibinfo
  {pages} {49} (\bibinfo {year} {2005})}\BibitemShut {NoStop}%
\bibitem [{\citenamefont {Brown}\ and\ \citenamefont
  {Gabrielse}(1986)}]{Brow1986}%
  \BibitemOpen
  \bibfield  {author} {\bibinfo {author} {\bibfnamefont {L.~S.}\ \bibnamefont
  {Brown}}\ and\ \bibinfo {author} {\bibfnamefont {G.}~\bibnamefont
  {Gabrielse}},\ }\href {\doibase 10.1103/RevModPhys.58.233} {\bibfield
  {journal} {\bibinfo  {journal} {Rev. Mod. Phys.}\ }\textbf {\bibinfo {volume}
  {58}},\ \bibinfo {pages} {233} (\bibinfo {year} {1986})}\BibitemShut
  {NoStop}%
\bibitem [{\citenamefont {Lohse}\ \emph {et~al.}(2019)\citenamefont {Lohse},
  \citenamefont {Berrocal}, \citenamefont {Block}, \citenamefont {Chemarev},
  \citenamefont {Cornejo}, \citenamefont {Ram\'irez},\ and\ \citenamefont
  {Rodr\'iguez}}]{Lohs2019}%
  \BibitemOpen
  \bibfield  {author} {\bibinfo {author} {\bibfnamefont {S.}~\bibnamefont
  {Lohse}}, \bibinfo {author} {\bibfnamefont {J.}~\bibnamefont {Berrocal}},
  \bibinfo {author} {\bibfnamefont {M.}~\bibnamefont {Block}}, \bibinfo
  {author} {\bibfnamefont {S.}~\bibnamefont {Chemarev}}, \bibinfo {author}
  {\bibfnamefont {J.~M.}\ \bibnamefont {Cornejo}}, \bibinfo {author}
  {\bibfnamefont {J.~G.}\ \bibnamefont {Ram\'irez}}, \ and\ \bibinfo {author}
  {\bibfnamefont {D.}~\bibnamefont {Rodr\'iguez}},\ }\href {\doibase
  10.1063/1.5094428} {\bibfield  {journal} {\bibinfo  {journal} {Rev. Sci.
  Instrum.}\ }\textbf {\bibinfo {volume} {90}},\ \bibinfo {pages} {063202}
  (\bibinfo {year} {2019})}\BibitemShut {NoStop}%
\bibitem [{\citenamefont {Comisarow}\ and\ \citenamefont
  {Marshall}(1974{\natexlab{a}})}]{Comi1974}%
  \BibitemOpen
  \bibfield  {author} {\bibinfo {author} {\bibfnamefont {M.~B.}\ \bibnamefont
  {Comisarow}}\ and\ \bibinfo {author} {\bibfnamefont {A.~G.}\ \bibnamefont
  {Marshall}},\ }\href {\doibase 10.1016/0009-2614(74)89137-2} {\bibfield
  {journal} {\bibinfo  {journal} {Chem.\ Phys.\ Lett.}\ }\textbf {\bibinfo
  {volume} {25}},\ \bibinfo {pages} {282} (\bibinfo {year}
  {1974}{\natexlab{a}})}\BibitemShut {NoStop}%
\bibitem [{\citenamefont {Comisarow}\ and\ \citenamefont
  {Marshall}(1974{\natexlab{b}})}]{Comi1974_2}%
  \BibitemOpen
  \bibfield  {author} {\bibinfo {author} {\bibfnamefont {M.~B.}\ \bibnamefont
  {Comisarow}}\ and\ \bibinfo {author} {\bibfnamefont {A.~G.}\ \bibnamefont
  {Marshall}},\ }\href {\doibase 10.1016/0009-2614(74)80397-0} {\bibfield
  {journal} {\bibinfo  {journal} {Chem.\ Phys.\ Lett.}\ }\textbf {\bibinfo
  {volume} {26}},\ \bibinfo {pages} {489} (\bibinfo {year}
  {1974}{\natexlab{b}})}\BibitemShut {NoStop}%
\bibitem [{\citenamefont {Lohse}\ \emph {et~al.}(2020)\citenamefont {Lohse},
  \citenamefont {Berrocal}, \citenamefont {B\"ohland}, \citenamefont {van~de
  Laar}, \citenamefont {Block}, \citenamefont {Chemarev}, \citenamefont
  {D\"ullman}, \citenamefont {Nagy}, \citenamefont {Ram\'irez},\ and\
  \citenamefont {Rodr\'iguez}}]{Lohs2020}%
  \BibitemOpen
  \bibfield  {author} {\bibinfo {author} {\bibfnamefont {S.}~\bibnamefont
  {Lohse}}, \bibinfo {author} {\bibfnamefont {J.}~\bibnamefont {Berrocal}},
  \bibinfo {author} {\bibfnamefont {S.}~\bibnamefont {B\"ohland}}, \bibinfo
  {author} {\bibfnamefont {J.}~\bibnamefont {van~de Laar}}, \bibinfo {author}
  {\bibfnamefont {M.}~\bibnamefont {Block}}, \bibinfo {author} {\bibfnamefont
  {S.}~\bibnamefont {Chemarev}}, \bibinfo {author} {\bibfnamefont {C.~E.}\
  \bibnamefont {D\"ullman}}, \bibinfo {author} {\bibfnamefont {S.}~\bibnamefont
  {Nagy}}, \bibinfo {author} {\bibfnamefont {J.~G.}\ \bibnamefont {Ram\'irez}},
  \ and\ \bibinfo {author} {\bibfnamefont {D.}~\bibnamefont {Rodr\'iguez}},\
  }\href {\doibase 10.1063/5.0015011} {\bibfield  {journal} {\bibinfo
  {journal} {Rev. Sci. Instrum.}\ }\textbf {\bibinfo {volume} {91}},\ \bibinfo
  {pages} {093202} (\bibinfo {year} {2020})}\BibitemShut {NoStop}%
\bibitem [{\citenamefont {Cornell}\ \emph {et~al.}(1990)\citenamefont
  {Cornell}, \citenamefont {Weisskoff}, \citenamefont {Boyce},\ and\
  \citenamefont {Pritchard}}]{Corn1990}%
  \BibitemOpen
  \bibfield  {author} {\bibinfo {author} {\bibfnamefont {E.~A.}\ \bibnamefont
  {Cornell}}, \bibinfo {author} {\bibfnamefont {R.~M.}\ \bibnamefont
  {Weisskoff}}, \bibinfo {author} {\bibfnamefont {K.~R.}\ \bibnamefont
  {Boyce}}, \ and\ \bibinfo {author} {\bibfnamefont {D.~E.}\ \bibnamefont
  {Pritchard}},\ }\href {\doibase 10.1103/PhysRevA.41.312} {\bibfield
  {journal} {\bibinfo  {journal} {Phys. Rev. A}\ }\textbf {\bibinfo {volume}
  {41}},\ \bibinfo {pages} {312} (\bibinfo {year} {1990})}\BibitemShut
  {NoStop}%
\bibitem [{\citenamefont {Giuliani}\ \emph {et~al.}(2019)\citenamefont
  {Giuliani}, \citenamefont {Matheson}, \citenamefont {Nazarewicz},
  \citenamefont {Olsen}, \citenamefont {Reinhard}, \citenamefont {Sadhukhan},
  \citenamefont {Schuetrumpf}, \citenamefont {Schunck},\ and\ \citenamefont
  {Schwerdtfeger}}]{Giul2019}%
  \BibitemOpen
  \bibfield  {author} {\bibinfo {author} {\bibfnamefont {S.~A.}\ \bibnamefont
  {Giuliani}}, \bibinfo {author} {\bibfnamefont {Z.}~\bibnamefont {Matheson}},
  \bibinfo {author} {\bibfnamefont {W.}~\bibnamefont {Nazarewicz}}, \bibinfo
  {author} {\bibfnamefont {E.}~\bibnamefont {Olsen}}, \bibinfo {author}
  {\bibfnamefont {P.-G.}\ \bibnamefont {Reinhard}}, \bibinfo {author}
  {\bibfnamefont {J.}~\bibnamefont {Sadhukhan}}, \bibinfo {author}
  {\bibfnamefont {B.}~\bibnamefont {Schuetrumpf}}, \bibinfo {author}
  {\bibfnamefont {N.}~\bibnamefont {Schunck}}, \ and\ \bibinfo {author}
  {\bibfnamefont {P.}~\bibnamefont {Schwerdtfeger}},\ }\href {\doibase
  10.1103/RevModPhys.91.011001} {\bibfield  {journal} {\bibinfo  {journal}
  {Rev.\ Mod.\ Phys.}\ }\textbf {\bibinfo {volume} {91}},\ \bibinfo {pages}
  {011001} (\bibinfo {year} {2019})}\BibitemShut {NoStop}%
\bibitem [{\citenamefont {Rodr\'iguez}\ \emph {et~al.}(2010)\citenamefont
  {Rodr\'iguez}, \citenamefont {Blaum}, \citenamefont {N\"ortersh\"auser},\
  and\ \citenamefont {\textit{et al.}}}]{Rodr2010}%
  \BibitemOpen
  \bibfield  {author} {\bibinfo {author} {\bibfnamefont {D.}~\bibnamefont
  {Rodr\'iguez}}, \bibinfo {author} {\bibfnamefont {K.}~\bibnamefont {Blaum}},
  \bibinfo {author} {\bibfnamefont {W.}~\bibnamefont {N\"ortersh\"auser}}, \
  and\ \bibinfo {author} {\bibnamefont {\textit{et al.}}},\ }\href {\doibase
  10.1140/epjst/e2010-01231-2} {\bibfield  {journal} {\bibinfo  {journal}
  {Eur.\ Phys.\ J.\ Special\ Topics}\ }\textbf {\bibinfo {volume} {183}},\
  \bibinfo {pages} {1} (\bibinfo {year} {2010})}\BibitemShut {NoStop}%
\bibitem [{\citenamefont {Hamaker}\ \emph {et~al.}(2019)\citenamefont
  {Hamaker}, \citenamefont {Bollen}, \citenamefont {Eibach}, \citenamefont
  {Izzo}, \citenamefont {Puentes}, \citenamefont {Redshaw}, \citenamefont
  {Ringle}, \citenamefont {Sandler}, \citenamefont {Schwarz},\ and\
  \citenamefont {Yandow}}]{Hama2019}%
  \BibitemOpen
  \bibfield  {author} {\bibinfo {author} {\bibfnamefont {A.}~\bibnamefont
  {Hamaker}}, \bibinfo {author} {\bibfnamefont {G.}~\bibnamefont {Bollen}},
  \bibinfo {author} {\bibfnamefont {M.}~\bibnamefont {Eibach}}, \bibinfo
  {author} {\bibfnamefont {C.}~\bibnamefont {Izzo}}, \bibinfo {author}
  {\bibfnamefont {D.}~\bibnamefont {Puentes}}, \bibinfo {author} {\bibfnamefont
  {M.}~\bibnamefont {Redshaw}}, \bibinfo {author} {\bibfnamefont
  {R.}~\bibnamefont {Ringle}}, \bibinfo {author} {\bibfnamefont
  {R.}~\bibnamefont {Sandler}}, \bibinfo {author} {\bibfnamefont
  {S.}~\bibnamefont {Schwarz}}, \ and\ \bibinfo {author} {\bibfnamefont
  {I.}~\bibnamefont {Yandow}},\ }\href {\doibase 10.1007/s10751-019-1576-9}
  {\bibfield  {journal} {\bibinfo  {journal} {Hyperfine Interact.}\ }\textbf
  {\bibinfo {volume} {240}} (\bibinfo {year} {2019}),\
  10.1007/s10751-019-1576-9}\BibitemShut {NoStop}%
\bibitem [{\citenamefont {Wineland}\ and\ \citenamefont
  {Dehmelt}(1975)}]{Wine1975}%
  \BibitemOpen
  \bibfield  {author} {\bibinfo {author} {\bibfnamefont {D.~J.}\ \bibnamefont
  {Wineland}}\ and\ \bibinfo {author} {\bibfnamefont {H.~G.}\ \bibnamefont
  {Dehmelt}},\ }\href {\doibase 10.1063/1.321602} {\bibfield  {journal}
  {\bibinfo  {journal} {J.\ Appl.\ Phys.}\ }\textbf {\bibinfo {volume} {46}},\
  \bibinfo {pages} {919} (\bibinfo {year} {1975})}\BibitemShut {NoStop}%
\bibitem [{\citenamefont {Guti{\'e}rrez}\ \emph {et~al.}(2019)\citenamefont
  {Guti{\'e}rrez}, \citenamefont {Berrocal}, \citenamefont {Cornejo},
  \citenamefont {Dom{\'i}nguez}, \citenamefont {Del~Pozo}, \citenamefont
  {Arrazola}, \citenamefont {Ba\~nuelos}, \citenamefont {Escobedo},
  \citenamefont {Block}, \citenamefont {Solano},\ and\ \citenamefont
  {Rodr{\'i}guez}}]{Guti2019}%
  \BibitemOpen
  \bibfield  {author} {\bibinfo {author} {\bibfnamefont {M.~J.}\ \bibnamefont
  {Guti{\'e}rrez}}, \bibinfo {author} {\bibfnamefont {J.}~\bibnamefont
  {Berrocal}}, \bibinfo {author} {\bibfnamefont {J.~M.}\ \bibnamefont
  {Cornejo}}, \bibinfo {author} {\bibfnamefont {F.}~\bibnamefont
  {Dom{\'i}nguez}}, \bibinfo {author} {\bibfnamefont {J.~J.}\ \bibnamefont
  {Del~Pozo}}, \bibinfo {author} {\bibfnamefont {I.}~\bibnamefont {Arrazola}},
  \bibinfo {author} {\bibfnamefont {J.}~\bibnamefont {Ba\~nuelos}}, \bibinfo
  {author} {\bibfnamefont {P.}~\bibnamefont {Escobedo}}, \bibinfo {author}
  {\bibfnamefont {M.}~\bibnamefont {Block}}, \bibinfo {author} {\bibfnamefont
  {E.}~\bibnamefont {Solano}}, \ and\ \bibinfo {author} {\bibfnamefont
  {D.}~\bibnamefont {Rodr{\'i}guez}},\ }\href {\doibase
  10.1088/1367-2630/aafa45} {\bibfield  {journal} {\bibinfo  {journal} {New J.
  Phys.}\ }\textbf {\bibinfo {volume} {21}},\ \bibinfo {pages} {023023}
  (\bibinfo {year} {2019})}\BibitemShut {NoStop}%
\bibitem [{\citenamefont {Smith~III}(2007)}]{Juli2007}%
  \BibitemOpen
  \bibfield  {author} {\bibinfo {author} {\bibfnamefont {J.~O.}\ \bibnamefont
  {Smith~III}},\ }\enquote {\bibinfo {title} {Mathematics of the discrete
  fourier transform (dft) with audio applications, second edition},}\ \
  (\bibinfo  {publisher} {W3K Publishing},\ \bibinfo {address} {Center for
  Computer Research in Music and Acoustics (CCRMA), Stanford University},\
  \bibinfo {year} {2007})\BibitemShut {NoStop}%
\bibitem [{\citenamefont {Kotler}\ \emph {et~al.}(2017)\citenamefont {Kotler},
  \citenamefont {Simmonds}, \citenamefont {Leibfried},\ and\ \citenamefont
  {Wineland}}]{Kotl2017}%
  \BibitemOpen
  \bibfield  {author} {\bibinfo {author} {\bibfnamefont {S.}~\bibnamefont
  {Kotler}}, \bibinfo {author} {\bibfnamefont {R.~W.}\ \bibnamefont
  {Simmonds}}, \bibinfo {author} {\bibfnamefont {D.}~\bibnamefont {Leibfried}},
  \ and\ \bibinfo {author} {\bibfnamefont {D.~J.}\ \bibnamefont {Wineland}},\
  }\href {\doibase 10.1103/PhysRevA.95.022327} {\bibfield  {journal} {\bibinfo
  {journal} {Phys. Rev. A}\ }\textbf {\bibinfo {volume} {95}},\ \bibinfo
  {pages} {022327} (\bibinfo {year} {2017})}\BibitemShut {NoStop}%
\end{thebibliography}
\providecommand{\noopsort}[1]{}\providecommand{\singleletter}[1]{#1}%
\end{document}